\documentclass[11pt]{article}
\usepackage[T1]{fontenc}
\usepackage[utf8]{inputenc}
\usepackage{lmodern,microtype}
\usepackage[margin=1in]{geometry}
\usepackage{amsmath,amssymb,amsthm,mathtools,mathrsfs}
\usepackage{booktabs,longtable,array,enumitem}
\usepackage[dvipsnames]{xcolor}
\usepackage[colorlinks=true,linkcolor=MidnightBlue,citecolor=MidnightBlue,urlcolor=MidnightBlue]{hyperref}
\usepackage{fancyhdr}
\hypersetup{bookmarksnumbered=true}
\allowdisplaybreaks[2]

\newtheorem{theorem}{Theorem}[section]
\newtheorem{lemma}[theorem]{Lemma}

\newtheorem{corollary}[theorem]{Corollary}
\newtheorem{theoremcorollary}{Corollary}[theorem]
\theoremstyle{remark}
\newcommand{\HH}{\mathcal H}
\newcommand{\BB}{\mathcal B}
\newcommand{\SSS}{\mathcal S}
\newcommand{\Se}{\mathcal S_e}

\newcommand{\one}{\mathbf1}
\newcommand{\dd}{\,\mathrm d}
\newcommand{\im}{\operatorname{Im}}
\newcommand{\re}{\operatorname{Re}}
\newcommand{\supp}{\operatorname{supp}}
\newcommand{\dist}{\operatorname{dist}}
\newcommand{\Dom}{\operatorname{Dom}}

\newcommand{\ip}[2]{\left\langle #1,#2\right\rangle}

\hypersetup{
  pdftitle={Dyson Equations: Existence and Estimates for the Self-Energy},
  pdfauthor={Yulia Karpeshina},
  pdfsubject={Existence and estimates for the self-energy}
}
\title{Dyson Equations:\\[3pt]
Existence and Estimates for the Self-Energy}
\author{Yulia Karpeshina\thanks{Partially supported by NSF Award No.~2605402.}\\[4pt]
{\small University of Alabama at Birmingham}\\
{\small\href{mailto:karpeshi@uab.edu}{\texttt{karpeshi@uab.edu}}}}
\date{}
\begin{document}
\maketitle
\begin{abstract}
We study a continuum Dyson equation in three-dimensional space in the
homogeneous case and for arbitrary measurable defect sets. We establish
existence and uniqueness of the solution, called the self-energy, in an
admissible class of closed operators on $L^2(\mathbb R^3)$. The solution
is bounded, with nonnegative imaginary part for $\im z>0$ and nonpositive
imaginary part for $\im z<0$, $z$ being the spectral parameter.
In each open half-plane, the self-energy depends holomorphically on $z$.
Its norm is uniformly bounded throughout both half-planes,
independently of the size and geometry of the defect set.
In the homogeneous case, the self-energy acts in momentum space as
multiplication by a scalar function $\mu(\xi,z)$. For sufficiently large
$k$ and $z=k^2+i\tau$ with $0<|\tau|\le1$, we prove
$\|\partial_\xi^\alpha\mu(\cdot,z)\|_\infty\le C_\alpha/k$
for every multi-index $\alpha$. We also obtain a positive lower bound
of order $k^{-1}$ for $\lvert\im\mu(\xi,z)\rvert$ in a fixed-width
neighborhood of the sphere $|\xi|=k$, uniformly in $\tau$.
\end{abstract}

\section{Introduction}\label{sec:introduction}\label{subsec:historical-introduction}
A self-consistent Dyson equation for the self-energy has the form
\begin{equation}\label{eq:intro-abstract-dyson}
 \Sigma(z)=\SSS\!\left[(H-zI-\Sigma(z))^{-1}\right],
\end{equation}
where $H$ is a given self-adjoint operator on a Hilbert space $\HH$,
$z$ is a complex spectral parameter, $\Sigma(z)$ is the unknown
operator, called the \emph{self-energy}, and $\SSS$ is a given bounded
linear map acting on the space of bounded operators on $\HH$.

Self-consistent Dyson equations provide a basic framework for describing propagation
and spectral properties in interacting and disordered quantum systems.
Their origins lie in the development of quantum electrodynamics,
including the work of F. J. Dyson~\cite{Dyson1949}, and they subsequently
became central to many-body theory and the theory of disordered media.
By incorporating scattering and interaction effects into a self-energy,
Dyson equations organize perturbation expansions and provide access to
averaged Green functions, spectral broadening, and the density of states;
see, for example, Akkermans and Montambaux \cite{AM}.

The mathematical study of self-consistent spectral equations has a
long history. Pastur \cite{Pastur1972} characterized limiting
random-matrix eigenvalue distributions through a nonlinear functional
equation. Helton, Rashidi Far, and Speicher \cite{HRS} established
existence and uniqueness for an operator-valued self-consistency equation
arising in random matrix theory. They also developed a convergent
iteration based on holomorphic fixed-point methods. Their work provides
an important foundation for subsequent analysis of Dyson equations.

Ajanki, Erd\H{o}s, and Kr\"uger \cite{AjEK} developed a quantitative
stability theory for the matrix Dyson equation and applied it to
random matrices when their size goes to infinity. Alt, Erd\H{o}s, and Kr\"uger \cite{AEK} studied the equation in
finite von Neumann algebras, which may be infinite dimensional, and
obtained detailed descriptions of spectral bands, edges, and cusps
under suitable hypotheses. Positivity, the imaginary-part identity,
and a balanced polar decomposition play important roles in this
analysis.

The present paper investigates a Dyson equation motivated by the study of
continuum Schr\"odinger operators with random potentials. Our results,
however, are entirely deterministic. We study
\begin{equation}\label{eq:intro-spatial-dyson}
 \Sigma_e(z)=\Se\!\left[
 (-\Delta+W-z-\Sigma_e(z))^{-1}
 \right],
\end{equation}
where $\Se$ combines the contributions of the translated
potentials $v(x-a)$. Namely, for a fixed integer
$\alpha_0\ge2$, we assume
\begin{equation}\label{eq:intro-single-site}
 v\in C_c^{\alpha_0}(\mathbb R^3),\qquad v(x)\ge0,\qquad v\not\equiv0.
\end{equation}
The translation parameter $a$ is called a \emph{scattering center},
or simply a \emph{center}. More precisely, for a bounded operator $T$
on $L^2(\mathbb R^3)$,
\begin{equation}\label{eq:intro-covariance}
 \Se[T]=\sigma^2\int_{\mathbb R^3\setminus \Omega}v_aTv_a\,\dd a,
 \qquad (v_af)(x)=v(x-a)f(x),
\end{equation}
with the fixed positive constant $\sigma^2$.
For each allowed center position, one multiplies $T$ on both sides
by the corresponding potential and integrates these contributions
over $a$. Each such potential retains its full spatial support,
which may extend across the boundary of $\Omega$.

\begin{samepage}
We present two main results:
\begin{enumerate}
\item \textbf{The exterior case:} $\Omega$ is an arbitrary measurable set,
including bounded sets and unbounded unions of deleted regions.
No separation, connectedness,
or boundary regularity is required. We assume
$W\in L^\infty_{\mathrm{loc}}(\mathbb R^3;\mathbb R)$ and
$W(x)\ge-C_W(1+|x|^2)$ almost everywhere for some finite $C_W\ge0$.
No support condition relating $W$ to $\Omega$ is needed.
\item \textbf{The homogeneous case:} $\Omega=\varnothing$ and $W=0$.
The self-energy is translation invariant and acts in momentum space as
multiplication by the scalar function $\mu(\xi,z)$ of Theorem~\ref{thm:mu}.
\end{enumerate}
\end{samepage}

In both settings, we construct a bounded self-energy on $L^2(\mathbb R^3)$
with nonnegative imaginary part for $\im z>0$ and nonpositive imaginary
part for $\im z<0$. We prove uniqueness in the admissible classes
specified in the main theorems, without assuming boundedness in advance.
The self-energy is holomorphic
in the spectral parameter $z$ in each open half-plane. Its operator norm
admits a uniform upper bound throughout both half-planes, with a constant
independent of the size and geometry of the defect set. In the homogeneous case, the self-energy
acts in momentum space as multiplication by a scalar function $\mu(\xi,z)$.
For sufficiently large $k$ and $z=k^2+i\tau$ with $0<|\tau|\le1$, we prove
$\|\partial_\xi^\alpha\mu(\cdot,z)\|_\infty\le C_\alpha/k$
for every multi-index $\alpha$. We also obtain a positive lower bound
of order $k^{-1}$ for $\lvert\im\mu(\xi,z)\rvert$ in a fixed-width
neighborhood of the sphere $|\xi|=k$, uniformly in $\tau$.

The homogeneous formulation is related to the self-consistent Born approximation
for random impurity scattering discussed by Horing
\cite[Section~5.4]{Horing}. Related translation-invariant self-consistent
equations for two-dimensional Dirac electrons appear in Hernando, Baba,
D\'iaz, and Dom\'inguez-Adame \cite[equations~(8a)--(8c)]{HBDD}.
These references concern the formulation and physical interpretation of the
self-consistent equation, not the uniform self-energy bounds proved here.

Our method follows the established approach to Dyson equations developed by
Ajanki--Erd\H{o}s--Kr\"uger~\cite{AjEK} and
Alt--Erd\H{o}s--Kr\"uger~\cite{AEK}. For existence and uniqueness, we use
the Earle--Hamilton fixed-point theorem~\cite{EH}, following the
operator-valued construction of Helton--Rashidi Far--Speicher~\cite{HRS}
underlying the existence theory in~\cite{AEK}. We also use the balanced
polar decomposition $R_e=QUQ^*$ and the transformed covariance map
$\Phi_e[X]=Q^*\Se[QXQ^*]Q$,~\cite{AEK,AjEK}. The imaginary-part identity gives
$\Phi_e[D]\le D$ for a nonnegative injective weight $D$. For each positive
covariance mode, a direct comparison then gives $BDB\le D$, where
$B=Q^*m_{e,h}Q\ge0$. Monotonicity of the square root implies $B\le I$.

There are essential differences in the setting. The earlier works concern
bounded matrices or bounded elements of a finite von Neumann algebra,
whereas $H_W=-\Delta+W$ is unbounded on $L^2(\mathbb R^3)$. The flatness
assumption used in their quantitative spectral analysis includes a
positive lower bound on the covariance map in terms of the normalized
trace. Such a global lower bound is unavailable here: the spatial
covariance can vanish on nonzero inputs supported sufficiently far
inside a deleted region. Moreover, $D$ need not have a bounded inverse.

The unbounded Hamiltonian requires the domain arguments in
Lemma~\ref{lem:exterior-factorization}; the direct order argument in
Section~\ref{sub:sandwich-mode-norm} uses the injectivity of $D$.
For detailed comparisons, see the paragraph following
\eqref{eq:self-consistency-im} and the remark on related results
following Lemma~\ref{lem:uniform-coefficient}.

The assumption $v\ge0$ is a technical hypothesis of the present approach
and can be weakened in the homogeneous high-energy analysis. We retain it
throughout for a unified treatment of the homogeneous and exterior problems.
The substantial uses of the sign assumption occur in
Sections~\ref{app:positive}--\ref{app:coef}, where we construct nonnegative
spatial weights and prove the uniform coefficient estimates. In
Section~\ref{app:hom}, there is an additional use through
$\widehat v(0)=\int v>0$.

The plan of the paper is the following. Section~\ref{sec:notation-assumptions}
introduces the notation, the covariance maps, and the assumptions on the
potentials and deleted center sets. Section~\ref{sec:main-results} presents
the exterior and homogeneous main results.
Section~\ref{sub:half-plane-reflection} explains how adjoint reflection
transfers the upper-half-plane results to the lower half-plane.

Section~\ref{sec:proofs} contains the proofs. In Section~\ref{app:exist},
we construct the self-energy by the holomorphic fixed-point method and
establish uniqueness, convergence of successive approximations, and
holomorphic dependence on the spectral parameter. Section~\ref{app:positive}
uses the balanced polar decomposition and a direct order argument
for positive covariance modes to obtain spatially weighted resolvent
bounds. These bounds are then used in
Section~\ref{app:coef} to prove the uniform operator-norm estimate for the
self-energy and its spatially weighted refinement. Section~\ref{app:hom}
develops the homogeneous high-energy analysis, including the bounds
$\|\partial_\xi^\alpha\mu\|_\infty\le C_\alpha/k$ for every multi-index
$\alpha$, the imaginary-part lower bound near the characteristic sphere,
and the boundedness and square-integrability estimates for the associated
resolvent multiplier.
Section~\ref{sub:technical-lemmas} collects the supporting technical lemmas:
automatic boundedness of admissible solutions and justification of the
resolvent factorization for an unbounded Hamiltonian.

Appendix~\ref{app:homogeneous-multiplier} derives the momentum representation
of the homogeneous covariance map. Appendix~\ref{app:transformed-covariance-identity}
verifies the transformed covariance identity used in Section~\ref{sub:saturated}.
Appendix~\ref{app:local-to-global} explains how the local estimates yield the
global operator bound for the self-energy.

\medskip
\noindent\textbf{Acknowledgements.}
The paper is made in collaboration with ChatGPT.
The author is thankful to Roman Shterenberg for useful discussions.

\section{Notation and assumptions}\label{sec:notation-assumptions}
Throughout, $\HH=L^2(\mathbb R^3)$, and $\BB=\BB(\HH)$ denotes
the space of bounded operators on $\HH$.
For spatial vectors $f\in\HH$, we consistently use
$\|f\|_{\HH}=\bigl(\int_{\mathbb R^3}|f(x)|^2\dd x\bigr)^{1/2}$.
For $T\in\BB(\HH)$, $\|T\|$ denotes its operator norm.
The $L^2$ norms in the center or momentum variable carry the space
explicitly, as in $\|h\|_{L^2(\Omega^c)}$ and
$\|r\|_{L^2(\mathbb R^3)}$, respectively.
For functions on $\mathbb R^3$, $\|\cdot\|_\infty$ denotes the
essential-supremum norm, in either the spatial or the momentum variable.
We denote the open upper and lower half-planes by
\begin{equation}
\mathbb C_\pm:=\{z\in\mathbb C:\pm\im z>0\}.
\end{equation}
We write the spectral parameter as
\begin{equation}
z=E+i\tau,\qquad E\in\mathbb R,\quad \tau\in\mathbb R\setminus\{0\}.
\end{equation}
In the high-energy estimates, we write $E=k^2$.

The notation $T\ge0$
means positivity of a self-adjoint operator; $\im T=(T-T^*)/(2i)$.
We denote by $B(x_0,r)$ the open ball of radius $r$ centered at $x_0$.
With the regularity index $\alpha_0$ from \eqref{eq:intro-single-site} and the constant
$\sigma^2$ from \eqref{eq:intro-covariance}, our standing assumptions are
\begin{equation}\label{eq:standing-single-site}
0\le v\in C_c^{\alpha_0}(\mathbb R^3),\quad v\not\equiv0,\quad
\supp v\subset B(0,s),\quad s>0,\quad \sigma>0.
\end{equation}

The multiplication operator associated with a center $a\in\mathbb R^3$
is denoted by $v_a$:
\begin{equation}\label{eq:single-site-translate}
(v_af)(x):=v(x-a)f(x),\qquad f\in\HH.
\end{equation}

\begin{samepage}
The constants in this article may depend on the fixed
function $v$, $\sigma^2$, and $s$; they do not depend on the defect or on $k,\tau$.
Throughout the homogeneous high-energy estimates, ``sufficiently large $k$''
means
\begin{equation}
k\ge k_0,\qquad k_0=k_0(v,\sigma^2,s)\ge1.
\end{equation}
\end{samepage}

The lower bound $k_0$ is independent of $\Omega$, $W$, and $\tau$.
It is also independent of the momentum and direction variables.
In particular, the choice of $k_0$ is unaffected by the sizes
or number of deleted regions or by the size of $W$.
The constant $k_0$ may be increased finitely many times in the proofs,
always with this dependence. When an auxiliary estimate involves an
arbitrary fixed width $H$, any additional dependence of its lower bound
on $H$ is stated explicitly. Some estimates stated for $k\ge1$
do not require the homogeneous high-energy restriction $k\ge k_0$.

For a fixed measurable set $A\subset\mathbb R^3$ of center points,
called a \emph{center set}, define the \emph{covariance kernel over $A$} by
\begin{equation}\label{eq:covariance-kernel-definition}
b_A(x,y)=\sigma^2\int_Av(x-a)v(y-a)\dd a.
\end{equation}
The associated \emph{covariance map over $A$} is defined by
\begin{equation}\label{eq:covariance-map-definition}
\SSS_A[T]=\sigma^2\int_A v_aTv_a\dd a.
\end{equation}
The operator integral is understood in the weak sense.

These are deterministic definitions: $v$, $\sigma^2$, and $A$ are
prescribed, and the integral is with respect to Lebesgue measure
on the center set $A$. No probability space or expectation is
used here. The term \emph{covariance} is the name for this kernel
and its associated map, motivated by the random-potential setting.
The set $A$ specifies which centers contribute;
it does not restrict the physical~points~$x,y$~to~$A$.

Let $\Omega\subset\mathbb R^3$ be an arbitrary measurable set of excluded
center points (a defect). The exterior covariance kernel $b_{\Omega^c}$ and covariance map $\SSS_{\Omega^c}$
are defined by \eqref{eq:covariance-kernel-definition} and \eqref{eq:covariance-map-definition}, respectively, with $A=\Omega^c$.
We denote the exterior covariance~map~by~$\Se$:
\begin{equation}\label{eq:exterior-covariance-map}
\Se[T]:=\SSS_{\Omega^c}[T]
=\sigma^2\int_{\Omega^c}v_aTv_a\dd a,\qquad T\in\BB(\HH).
\end{equation}
We also write $\SSS=\SSS_{\mathbb R^3}$ for the homogeneous covariance
map over all centers and $\SSS_\Omega$ for the covariance map over centers
in $\Omega$. The corresponding kernels are $b_{\mathbb R^3}$,
$b_{\Omega^c}$, and $b_\Omega$. In particular
$\SSS=\Se+\SSS_\Omega$ and
$b_{\mathbb R^3}=b_{\Omega^c}+b_\Omega$.

In the exterior covariance
map $\Se=\SSS_{\Omega^c}$, the integral is taken over center points
in $\Omega^c$, while the physical variables $x,y$ range over all of
$\mathbb R^3$. Consequently, the kernel $b_{\Omega^c}(x,y)$ may be
nonzero even when $x$ or $y$ lies inside $\Omega$ near its boundary.

For a set $A\subset\mathbb R^3$ and a number $t>0$, define its
\emph{open $t$-neighborhood} by
\begin{equation}
A^{+t}:=\{x\in\mathbb R^3:\dist(x,A)<t\},\qquad
\dist(x,A):=\inf_{a\in A}|x-a|.
\end{equation}
We use $\dist(x,\varnothing)=+\infty$, so
$\varnothing^{+t}=\varnothing$.

For the exterior problem, we assume that $W$ is real and locally
essentially bounded, and that there is a finite constant $C_W\ge0$ such that
\begin{equation}
W(x)\ge-C_W(1+|x|^2)\quad\text{for almost every }x\in\mathbb R^3.
\end{equation}
Define $H_W$ as the closure of $-\Delta+W$ initially acting on
$C_c^\infty(\mathbb R^3)$. This closure is self-adjoint by the
Faris--Lavine theorem; see Reed--Simon
\cite[Section~X.5, Theorem~X.38 and its following corollaries,
pp.~198--200]{RSII}. The adjoint-domain characterization and local
elliptic regularity give
\begin{equation}\label{eq:countable-local-domain}
\Dom H_W=\{u\in\HH:-\Delta u+Wu\in\HH
\text{ in distributions}\}\subset H^2_{\mathrm{loc}}(\mathbb R^3).
\end{equation}
All inverse operators below are bounded inverses of closed operators
with domain $\Dom H_W$, unless a different space is explicitly specified.

We denote the unitary Fourier transform from $L^2(\mathbb R^3_x)$
to $L^2(\mathbb R^3_\xi)$ by $\mathcal F$, but write
$\widehat v(\xi)=\int e^{-i\xi\cdot x}v(x)\dd x$ without normalization.
The symbol $*$ denotes convolution in momentum space. For a function $u$,
$M_u$ means multiplication by $u$ in the specified variable:
$(M_uf)(x)=u(x)f(x)$ in physical space, and
$(M_uf)(\xi)=u(\xi)f(\xi)$ in momentum space.
We use $\chi$ for smooth cutoff functions. For a measurable set $A$,
$\one_A$ denotes its characteristic function and the corresponding
multiplication operator.

In Theorem~\ref{thm:mu} we will show that the
homogeneous self-energy $\Sigma_0(z)$\footnote{The subscript $0$ denotes the homogeneous case
$\Omega=\varnothing$, $W=0$; we retain $\mu$ for the scalar function of momentum.}
is translation invariant and acts in momentum space by multiplication by
a scalar function $\mu(\xi,z)$. In this representation, the Dyson
equation \eqref{eq:intro-spatial-dyson} takes the scalar form
\eqref{eq:mu}. We define the corresponding resolvent multiplier in
momentum space by
\begin{equation}\label{eq:hom-resolvent-multiplier}
r(\xi,z):=\frac1{|\xi|^2-z-\mu(\xi,z)}.
\end{equation}

We define the exterior coverage function $m_e$ and the total mass $m$ by
\begin{equation}\label{eq:exterior-coverage}
0\le m_e(x):=\int_{\Omega^c}v(x-a)\dd a
\le m:=\int_{\mathbb R^3}v(a)\dd a.
\end{equation}
Here $m>0$ because $v$ is nonnegative and not identically zero;
the upper bound follows by enlarging the integration domain to
$\mathbb R^3$. The function $m_e$ measures the single-site weight
supplied by centers outside $\Omega$ at the physical point $x$.

For a measurable spatial set $U$, define the localized profile by
\begin{equation}\label{eq:localized-coverage}
m_{e,U}(x)=\one_U(x)m_e(x).
\end{equation}
We use the same symbol for a nonnegative profile and its multiplication
operator.

For each real, nonnegative $h\in L^2(\Omega^c)$ with $\|h\|_{L^2(\Omega^c)}=1$,
define the multiplication operator $m_{e,h}$ by
\begin{equation}\label{eq:mode-multiplier}
 m_{e,h}(x)=\sigma\int_{\Omega^c}h(a)v(x-a)\dd a,
 \qquad (m_{e,h}f)(x)=m_{e,h}(x)f(x).
\end{equation}
By Cauchy--Schwarz, $0\le m_{e,h}(x)\le\sigma\|v\|_{\HH}$.
We call $m_{e,h}$ a \emph{positive mode of the covariance}, or simply a \emph{mode}.
Here $h$ weights the center variable $a\in\Omega^c$, whereas $m_{e,U}$ is
the restriction of the exterior profile $m_e$ to the spatial set $U$.

The unit cubes are
$C_p=p+[-1/2,1/2)^3$, $p\in\mathbb Z^3$.
A unit cube $C_p$ will be called a \emph{bath cube}
if $C_p\cap\Omega^{+s}=\varnothing$.

\newpage
\section{Main results}\label{sec:main-results}
\subsection{The exterior self-energy for an arbitrary measurable deleted set}
\begin{theorem}[General deleted sets]\label{thm:general-exterior}
Assume the single-site hypotheses \eqref{eq:standing-single-site}. Let $\Omega\subset\mathbb R^3$ be
measurable and let $W\in L^\infty_{\mathrm{loc}}(\mathbb R^3)$ be real,
with
\begin{equation}
W(x)\ge-C_W(1+|x|^2)\quad\text{a.e.},\qquad C_W<\infty.
\end{equation}
Let $H_W$ be the self-adjoint closure specified in
Section~\ref{sec:notation-assumptions}.

\textup{\textbf{1.}}~For every $z\in\mathbb C\setminus\mathbb R$,
the equation
\begin{equation}\label{eq:general-dyson}
\Sigma_e(z)=\SSS_{\Omega^c}[R_e(z)],\qquad
R_e(z)=(H_W-z-\Sigma_e(z))^{-1}
\end{equation}
has a unique solution in the class of bounded operators satisfying
$\im\Sigma_e(z)\ge0$ for $z\in\mathbb C_+$ and
$\im\Sigma_e(z)\le0$ for $z\in\mathbb C_-$.

\textup{\textbf{2.}}~For measurable spatial sets $U_1,U_2$,
\begin{equation}\label{eq:general-spatial-support}
\one_{U_1}\Sigma_e(z)\one_{U_2}=0
\quad\text{if }\dist(U_1,U_2)>2s.
\end{equation}

\textup{\textbf{3.}}~Both $\Sigma_e(z)$ and $R_e(z)$ are holomorphic in operator norm in each
open half-plane. For every $z\in\mathbb C\setminus\mathbb R$, we have
\begin{equation}\label{eq:general-coefficient-bound}
\|\Sigma_e(z)\|\le C_v,
\end{equation}
and for every $u\in\HH$,
\begin{equation}\label{eq:K-coverage}
\|\Sigma_e(z)u\|_{\HH}+\|\Sigma_e(z)^*u\|_{\HH}
\le C_v\|u\|_{\HH}^{5/7}\|m_e^{1/2}u\|_{\HH}^{2/7}.
\end{equation}
Here the constant $C_v$ depends only on $v,\sigma^2,s$
(it is independent of $\Omega,W,z$).
For $R=R_e(z)$ or $R=R_e(z)^*$, all real, nonnegative unit vectors
$h_0,h_1\in L^2(\Omega^c)$, and the modes $m_{e,h_i}$, $i=0,1$, defined by
\eqref{eq:mode-multiplier}, the following estimate holds
\begin{equation}\label{eq:K-mixed-mode}
\|m_{e,h_0}^{1/2}R m_{e,h_1}^{1/2}\|_{L^2\to L^2}\le1.
\end{equation}

\begin{samepage}
\textup{\textbf{4.}}~There is uniqueness in a wide class of solutions.
Namely, fix $z\in\mathbb C\setminus\mathbb R$ and let
$\widetilde\Sigma$ be a closed, densely defined operator on
$\HH=L^2(\mathbb R^3)$ such that
\begin{equation}
\operatorname{sgn}(\im z)\,\im\langle\widetilde\Sigma u,u\rangle\ge0,
\qquad u\in\Dom\widetilde\Sigma.
\end{equation}
Assume that the common domain
$\mathcal D=\Dom H_W\cap\Dom\widetilde\Sigma$ is dense and that
$H_W-z-\widetilde\Sigma:\mathcal D\to\HH$ is bijective.
Denote its everywhere-defined inverse by $\widetilde R$, and
assume that the Dyson equation holds in the weak sense
\begin{equation}
\langle\widetilde\Sigma u,f\rangle
=\sigma^2\int_{\Omega^c}\langle\widetilde Rv_au,v_af\rangle\dd a,
\qquad u\in\Dom\widetilde\Sigma,\quad f\in\HH.
\end{equation}
Then $\Dom\widetilde\Sigma=\HH$,
$\widetilde\Sigma=\Sigma_e(z)$, and $\widetilde R=R_e(z)$.\footnote{Neither $\widetilde\Sigma$ nor $\widetilde R$ is assumed bounded
in advance; boundedness follows from these conditions.}
\end{samepage}

\textup{\textbf{5.}}~For every fixed $z\in\mathbb C\setminus\mathbb R$, the successive
approximations
\begin{equation}\label{eq:iteration}
\Sigma_e^{(0)}(z)=0,\qquad
\Sigma_e^{(n+1)}(z)=\SSS_{\Omega^c}\!\left[
(H_W-z-\Sigma_e^{(n)}(z))^{-1}\right],
\end{equation}
converge in operator norm to $\Sigma_e(z)$.
\end{theorem}

\noindent{\em Remark.}
No boundedness, separation, connectedness, or boundary regularity of $\Omega$
is required. No relation between $\Omega$ and the support of $W$ is needed
for these conclusions. In particular, the theorem applies to a bounded
measurable $\Omega$ and a bounded real $W$ that vanishes almost everywhere
outside $\Omega^{+s}$.

\begin{theoremcorollary}[Vanishing inside the deleted set]\label{cor:interior-vanishing}
Under the hypotheses of Theorem~\ref{thm:general-exterior}, let $U\subset\Omega$
be measurable and suppose that $\dist(U,\Omega^c)>s$. Then
\begin{equation}
\Sigma_e(z)\one_U=\one_U\Sigma_e(z)=0,\qquad z\in\mathbb C\setminus\mathbb R.
\end{equation}
\end{theoremcorollary}
\begin{proof}
If $U\subset\Omega$ is measurable and $\dist(U,\Omega^c)>s$, then
$m_e=0$ on $U$ by \eqref{eq:exterior-coverage}. Applying \eqref{eq:K-coverage}
to $u=\one_U f$ and taking adjoints gives
$\Sigma_e(z)\one_U=\one_U\Sigma_e(z)=0$.
\end{proof}

\begin{proof}[Proof of Theorem~\ref{thm:general-exterior}]
It suffices to work in $\mathbb C_+$, by Section~\ref{sub:half-plane-reflection}.
Sections~\ref{sub:covariance-properties} and~\ref{sub:fixed-point-construction}
establish existence and uniqueness of $\Sigma_e(z)$ among bounded operators
with nonnegative imaginary part and convergence of the iteration
\eqref{eq:iteration} in operator norm. Section~\ref{sub:holomorphic-construction}
proves that $\Sigma_e(z)$ and $R_e(z)$ depend holomorphically on $z$.
The finite-range property \eqref{eq:general-spatial-support} is given by
Corollary~\ref{cor:exterior-finite-range}. Section~\ref{app:coef} proves the
uniform estimates \eqref{eq:general-coefficient-bound} and \eqref{eq:K-coverage},
and Section~\ref{app:positive} gives \eqref{eq:K-mixed-mode}.
Uniqueness in the wider class of admissible closed operators follows from
Lemma~\ref{prop:admissible-exterior}.
\end{proof}

\paragraph{Scope of the exterior theorem.}
For $z=E+i\tau$ with $\tau\ne0$, the imaginary-part sign gives
$\|R_e(z)\|\le1/|\tau|$. The bound on $\Sigma_e$ is uniform as
$|\tau|\downarrow0$.
Theorem~\ref{thm:general-exterior} does not assert a uniform
unweighted resolvent bound, a real-axis boundary limit, or a rate of
iteration convergence uniform down to $\tau=0$.

\subsection{The homogeneous self-energy}
\begin{theorem}\label{thm:mu}
\textup{\textbf{1.}}~For every $z\in\mathbb C\setminus\mathbb R$, the homogeneous Dyson equation
\begin{equation}\label{eq:hom-dyson-operator}
\Sigma_0(z)=\SSS[R_\mu(z)],\qquad
R_\mu(z)=(-\Delta-z-\Sigma_0(z))^{-1}
\end{equation}
has a unique solution $\Sigma_0(z)$ in the class of bounded operators
on $\HH$ satisfying $\operatorname{sgn}(\im z)\,\im\Sigma_0(z)\ge0$.
For all measurable sets $U_1,U_2\subset\mathbb R^3$, this solution satisfies
\begin{equation}\label{eq:mu-finite-range}
\one_{U_1}\Sigma_0(z)\one_{U_2}=0
\qquad\text{if }\dist(U_1,U_2)>2s.
\end{equation}

\textup{\textbf{2.}}~The solution $\Sigma_0(z)$ is translation invariant. Let
$\mu(\cdot,z)\in L^\infty(\mathbb R^3)$ be its momentum multiplier, so that
$\widehat{\Sigma_0(z)u}(\xi)=\mu(\xi,z)\widehat u(\xi)$ for $u\in\HH$.
Under this correspondence, the operator equation \eqref{eq:hom-dyson-operator}
is equivalent to the scalar equation
\begin{equation}\label{eq:mu}
\mu(\xi,z)=\frac{\sigma^2}{(2\pi)^3}\int_{\mathbb R^3}
\frac{|\widehat v(\xi-\eta)|^2}{|\eta|^2-z-\mu(\eta,z)}\dd\eta,
\end{equation}
with the solution $\mu(\xi,z)$ satisfying
$\operatorname{sgn}(\im z)\,\im\mu(\xi,z)\ge0$ for every $\xi\in\mathbb R^3$.
The finite-range property \eqref{eq:mu-finite-range} is equivalent to
\begin{equation}\label{eq:mu-spatial-support}
\supp\bigl(\mathcal F^{-1}_{\xi\to x}\mu(\cdot,z)\bigr)
\subseteq\overline{B(0,2s)},\qquad z\in\mathbb C\setminus\mathbb R.
\end{equation}

\textup{\textbf{3.}}~The scalar solution of \eqref{eq:mu} is holomorphic in $z$ in each open half-plane.
For every multi-index $\alpha\in\mathbb N_0^3$, there is a constant
$C_{\alpha,v}$, depending only on $\alpha,v,\sigma^2$, and $s$, such that
\begin{equation}\label{eq:mu-global-derivatives}
\|\partial_\xi^\alpha\mu(\cdot,z)\|_\infty\le C_{\alpha,v},
\qquad z\in\mathbb C\setminus\mathbb R.
\end{equation}
There exists $k_0\ge1$, depending only on $v,\sigma^2,s$
(independent of $\Omega,W,\tau$), such that for every $k\ge k_0$,
uniformly for $0<|\tau|\le1$,
\begin{equation}\label{eq:mu-bounds}
\|\partial_\xi^\alpha\mu(\cdot,k^2+i\tau)\|_\infty
\le C_\alpha/k
\end{equation}
for every fixed multi-index $\alpha$. The constant $C_\alpha$ may
also depend on $\alpha$, but the lower bound $k_0$ does not.

\textup{\textbf{4.}}~The homogeneous resolvent multiplier $r$ defined by
\eqref{eq:hom-resolvent-multiplier} satisfies, in the same range of $k$ and $\tau$,
\begin{equation}\label{eq:r-bounds}
\|r(\cdot,k^2+i\tau)\|_\infty\le Ck,
\qquad
\|r(\cdot,k^2+i\tau)\|_{L^2(\mathbb R^3)}^2\le Ck^2.
\end{equation}

\textup{\textbf{5.}}~There are fixed $h,c,C>0$, depending only on $v,s,\sigma$, such that in the same range of $k,\tau$,
\begin{equation}\label{eq:mu-shell-bounds}
\frac{c}{k}\le\lvert\im\mu(\xi,k^2+i\tau)\rvert\le\frac{C}{k},
\qquad \text{when }\bigl|\,|\xi|-k\,\bigr|\le h.
\end{equation}
More precisely, uniformly in $\xi\in\mathbb R^3$ and $0<|\tau|\le1$,
\begin{equation}\label{eq:mu-shell-asymptotic}
\lvert\im\mu(\xi,k^2+i\tau)\rvert
=\frac{\sigma^2}{16\pi^2 k}
\int_{|\eta|=k}|\widehat v(\xi-\eta)|^2\dd S(\eta)+O(k^{-2}).
\end{equation}

\textup{\textbf{6.}}~For each fixed $z\in\mathbb C\setminus\mathbb R$, the scalar solution is unique
in the wider class of measurable functions
$\widetilde\mu:\mathbb R^3\to\mathbb C$ that are finite almost
everywhere, satisfy $\operatorname{sgn}(\im z)\,\im\widetilde\mu\ge0$
almost everywhere, and solve \eqref{eq:mu} almost everywhere. Every such solution equals
$\mu(\cdot,z)$ almost everywhere. No boundedness or regularity of
$\widetilde\mu$ is assumed in this uniqueness statement.

\begin{samepage}
\textup{\textbf{7.}}~The iteration
\begin{equation}
\mu_0=0,\qquad
\mu_{n+1}=\frac{\sigma^2}{(2\pi)^3}|\widehat v|^2*(|\cdot|^2-z-\mu_n)^{-1}
\end{equation}
converges to $\mu(\cdot,z)$ in $L^\infty(\mathbb R^3)$ for each fixed $z\in\mathbb C\setminus\mathbb R$.
The corresponding operator iterates converge to $\Sigma_0(z)$ in the
operator norm of $\BB(\HH)$ and are given by
\begin{equation}
\Sigma_0^{(0)}(z)=0,\qquad
\Sigma_0^{(n+1)}(z)=\SSS\!\left[(-\Delta-z-\Sigma_0^{(n)}(z))^{-1}\right].
\end{equation}
\end{samepage}
\end{theorem}

\begin{theoremcorollary}[Ward identities]\label{cor:homogeneous-ward}
Let $\mu$ be the solution of Theorem~\ref{thm:mu}, and let $r$
be defined by \eqref{eq:hom-resolvent-multiplier}.
For $z\in\mathbb C\setminus\mathbb R$, set $\tau=\im z$ and
$\gamma(\xi,z)=\tau+\im\mu(\xi,z)$. Then
\begin{equation}\label{eq:ward-hom}
\im r=\gamma|r|^2,\qquad \gamma=\tau+\frac{\sigma^2}{(2\pi)^3}|\widehat v|^2*(\gamma|r|^2).
\end{equation}
\end{theoremcorollary}
\begin{proof}
Set $a(\xi,z)=|\xi|^2-\re z-\re\mu(\xi,z)$.
By \eqref{eq:hom-resolvent-multiplier}, $r=(a-i\gamma)^{-1}$,
which gives the first identity in \eqref{eq:ward-hom}.
Taking imaginary parts in \eqref{eq:mu} and using the first identity
gives the second.
\end{proof}

\begin{proof}[Proof of Theorem~\ref{thm:mu}]
It suffices to work in $\mathbb C_+$ by the reflection argument of
Section~\ref{sub:half-plane-reflection}.
Step~6 of Section~\ref{sub:fixed-point-construction}, applied with
$H=-\Delta$ and $A=\mathbb R^3$, gives a unique bounded solution
$\Sigma_0(z)$ of \eqref{eq:hom-dyson-operator} with $\im\Sigma_0(z)\ge0$.

The finite-range property \eqref{eq:mu-finite-range} is given by Corollary~\ref{cor:exterior-finite-range}.

Let $T_bu(x)=u(x-b)$ for $b\in\mathbb R^3$. Since
$T_bv_aT_b^*=v_{a+b}$, a change of the center variable in
\eqref{eq:covariance-map-definition} gives
$T_b\SSS[X]T_b^*=\SSS[T_bXT_b^*]$. The free Laplacian commutes with
$T_b$, so $T_b\Sigma_0(z)T_b^*$ solves \eqref{eq:hom-dyson-operator}
with nonnegative imaginary part. Uniqueness gives
$T_b\Sigma_0(z)T_b^*=\Sigma_0(z)$ for every $b$.

Hence, a bounded translation-invariant operator $\Sigma_0(z)$ with nonnegative imaginary
part has the form $\Sigma_0(z)=\mathcal F^{-1}M_{\mu(\cdot,z)}\mathcal F$, where
$\mu(\cdot,z)\in L^\infty(\mathbb R^3)$ and $\im\mu(\cdot,z)\ge0$ almost
everywhere. The resolvent $(-\Delta-z-\Sigma_0(z))^{-1}$ has momentum multiplier
$r(\xi,z)=(|\xi|^2-z-\mu(\xi,z))^{-1}$ as in
\eqref{eq:hom-resolvent-multiplier}, with $\|r(\cdot,z)\|_\infty\le1/\im z$.
By \eqref{eq:momentum-convolution} with $f=r$, the operator
$\SSS[(-\Delta-z-\Sigma_0(z))^{-1}]$ is Fourier equivalent to
$\frac{\sigma^2}{(2\pi)^3}|\widehat v|^2*r(\cdot,z)$.
Thus $\Sigma_0(z)$ solves \eqref{eq:hom-dyson-operator} if and only if its multiplier
$\mu(\cdot,z)$ solves \eqref{eq:mu}.
By Fourier inversion in \eqref{eq:homogeneous-limit-operator},
$\mathcal F^{-1}_{\xi\to x}\mu(\cdot,z)$ has the same support as the
convolution kernel of $\Sigma_0(z)$. The finite-range property
\eqref{eq:mu-finite-range} therefore gives \eqref{eq:mu-spatial-support}.

Section~\ref{sub:holomorphic-construction} shows that $\mu_n(\cdot,z)$ is the
momentum multiplier of $\Sigma_0^{(n)}(z)$. By \eqref{eq:local-uniform-iteration}, \eqref{eq:homogeneous-iterate-norm-identity}, \eqref{eq:homogeneous-scalar-convergence},
$\|\Sigma_0^{(n)}(z)-\Sigma_0(z)\|_{\BB(\HH)}
=\|\mu_n(\cdot,z)-\mu(\cdot,z)\|_\infty\to0$, proving both convergence
statements in part~7. The same section establishes holomorphy of
$z\mapsto\mu(\cdot,z)$ in $\mathbb C_+$.

Applying Lemma~\ref{lem:uniform-coefficient} with $\Omega=\varnothing$ and
$W=0$ gives $\|\mu(\cdot,z)\|_\infty=\|\Sigma_0(z)\|\le C_v$.
By \eqref{eq:mu-spatial-support}, $\mu=\psi*\mu$ with
$\psi=(2\pi)^{-3}\widehat\chi$, where $\chi\in C_c^\infty(\mathbb R^3)$
depends only on $s$ and equals one near $\overline{B(0,2s)}$.
Differentiating the convolution gives
$\|\partial_\xi^\alpha\mu\|_\infty\le\|\partial^\alpha\psi\|_1\|\mu\|_\infty
\le C_v\|\partial^\alpha\psi\|_1$, proving \eqref{eq:mu-global-derivatives}.

Section~\ref{app:hom} derives all the high-energy estimates.
Formula \eqref{eq:mu-shell-asymptotic} follows from
\eqref{eq:hom-im-asymptotic} and reflection. Step~6 of
Lemma~\ref{lem:hom-shell-lower} gives the lower bound in
\eqref{eq:mu-shell-bounds}. The upper bound follows from \eqref{eq:mu-bounds} with $\alpha=0$.
The $L^\infty$ and $L^2$ estimates in \eqref{eq:r-bounds}
follow from Lemmas~\ref{lem:hom-shell-lower}
and~\ref{lem:hom-resolvent-l2}, respectively.

Finally, any measurable solution $\widetilde\mu$ in the stated
class is bounded by Lemma~\ref{prop:admissible-mu}. The equivalence
just proved makes $\mathcal F^{-1}M_{\widetilde\mu}\mathcal F$ a bounded
operator solution with the same sign. Hence it equals $\Sigma_0(z)$, so
$\widetilde\mu=\mu(\cdot,z)$ almost everywhere.
\end{proof}

\subsection{Adjoint reflection and the lower half-plane}
\label{sub:half-plane-reflection}
The lower-half-plane assertions in Theorems~\ref{thm:general-exterior}--\ref{thm:mu}
and Corollary~\ref{cor:homogeneous-ward}
follow from their upper-half-plane counterparts by adjoint reflection.
For $z\in\mathbb C_-$, define
\begin{equation}\label{eq:adjoint-reflection}
\Sigma_e(z):=\Sigma_e(\overline z)^*,\qquad
R_e(z):=R_e(\overline z)^*,\qquad
\mu(\xi,z):=\overline{\mu(\xi,\overline z)},\qquad
\Sigma_0(z)=\Sigma_0(\overline z)^*.
\end{equation}
Here the values on the right are those constructed in $\mathbb C_+$.
Self-adjointness of $H_W$ and the identity
$\SSS_A[X^*]=\SSS_A[X]^*$, which follows from the real-valuedness of $v$,
give
\begin{equation}
R_e(\overline z)^*
=(H_W-z-\Sigma_e(\overline z)^*)^{-1},\qquad
\Sigma_e(\overline z)^*=\Se[R_e(\overline z)^*].
\end{equation}
Thus the reflected operators satisfy the same Dyson equation.
Their self-energy has nonpositive imaginary part because
$\im\Sigma_e(z)=-\im\Sigma_e(\overline z)$.
The reflection is holomorphic in $\mathbb C_-$: a norm-convergent local
power series $\sum_{n\ge0}A_n(w-w_0)^n$ is transformed into
$\sum_{n\ge0}A_n^*(z-\overline{w_0})^n$.
Taking adjoints also preserves norms and transfers the fixed-point
iteration and uniqueness among bounded solutions to $\mathbb C_-$.

For the wider uniqueness classes, the same automatic-boundedness
argument applies with $|\im z|$ in place of $\im z$.
For a measurable scalar solution with the prescribed sign,
\mbox{$\bigl||\eta|^2-z-\widetilde\mu(\eta)\bigr|\ge|\im z|$}, so
Lemma~\ref{prop:admissible-mu} applies with this replacement.
For a closed operator coefficient with
$\operatorname{sgn}(\im z)\,\im\langle\widetilde\Sigma u,u\rangle\ge0$,
one has
\begin{equation}
\|(H_W-z-\widetilde\Sigma)u\|_{\HH}\ge|\im z|\,\|u\|_{\HH},
\qquad u\in\Dom H_W\cap\Dom\widetilde\Sigma.
\end{equation}
Hence $\|\widetilde R\|\le|\im z|^{-1}$. The weak Dyson identity and
closedness then imply $\Dom\widetilde\Sigma=\HH$ and
$\|\widetilde\Sigma\|\le\sigma^2\|v\|_{\HH}^2/|\im z|$, exactly as in
Lemma~\ref{prop:admissible-exterior}. Uniqueness now reduces to the
bounded-solution case above.

The spatial support and weighted estimates also transfer under adjoints;
in the two-mode estimate the two modes are interchanged. All constants
are unchanged. In the homogeneous case,
$r(\xi,z)=\overline{r(\xi,\overline z)}$, so the estimates
\eqref{eq:mu-bounds}--\eqref{eq:mu-shell-asymptotic} hold for
$0<|\tau|\le1$, with $|\im\mu|$ in
\eqref{eq:mu-shell-bounds} and \eqref{eq:mu-shell-asymptotic}. The Ward identities \eqref{eq:ward-hom}
remain valid with the signed quantities $\tau$ and $\gamma$.
This construction defines the solutions on the two open half-planes;
it does not assert a boundary limit or analytic continuation through
the real axis.

\section{Proofs}\label{sec:proofs}
\numberwithin{theorem}{subsection}

\subsection{Existence, uniqueness and holomorphic dependence}\label{app:exist}
\subsubsection{Boundedness and positivity of the covariance map}
\label{sub:covariance-properties}
The fixed point construction below uses boundedness, complex linearity,
and positivity of the covariance map. For every measurable center set
$A$, consider $\SSS_A$ defined in \eqref{eq:covariance-map-definition},
acting on $\BB(\HH)$ with $\HH=L^2(\mathbb R^3)$.
Here $v_a$ is the multiplication operator in
\eqref{eq:single-site-translate}. The integral definition shows
that $\SSS_A$ is complex-linear.

For $T\in\BB(\HH)$ and $f,g\in\HH$,
\begin{equation}
|\ip f{\SSS_A[T]g}|
\le \sigma^2\|T\|
\left(\int_A\|v_af\|_{\HH}^2\dd a\right)^{1/2}
\left(\int_A\|v_ag\|_{\HH}^2\dd a\right)^{1/2}
\le \sigma^2\|v\|_{\HH}^2\|T\|\|f\|_{\HH}\|g\|_{\HH}.
\end{equation}
Thus $\SSS_A$ is bounded, with $\|\SSS_A\|\le\sigma^2\|v\|_{\HH}^2$.

An operator $X$ is \emph{nonnegative}, written $X\ge0$, if it is
self-adjoint and $\langle f,Xf\rangle\ge0$ for every $f\in\HH$.
A linear map is \emph{positive} if it sends nonnegative operators
to nonnegative operators.
If $T\ge0$, self-adjointness of $v_a$ gives
\begin{equation}
\langle f,\SSS_A[T]f\rangle
=\sigma^2\int_A\langle v_af,Tv_af\rangle\dd a\ge0
\qquad(f\in\HH).
\end{equation}
This proves positivity of $\SSS_A$.

\subsubsection{The fixed point domain and the unbounded differential operator}\label{sub:fixed-point-construction}

\paragraph{Meaning of holomorphy in this section.}
Let $\mathcal U$ be an open subset of a complex Banach space
$\mathcal X$. A map $F:\mathcal U\to\mathcal X$ is
\emph{holomorphic} if it is complex Fr\'echet differentiable at $B$
whenever $B$ belongs to $\mathcal U$. This means that there is a bounded
complex-linear map $DF(B):\mathcal X\to\mathcal X$ such that
\begin{equation}
\lim_{\substack{\|C\|\to0\\ C\ne0}}
\frac{\|F(B+C)-F(B)-DF(B)[C]\|}{\|C\|}=0.
\end{equation}
The increment $C$ is arbitrary, subject to $B+C\in\mathcal U$;
openness ensures this for every sufficiently small $C$.
The notation $DF(B)[C]$ means the derivative at the point $B$
applied to the increment $C$.
Boundedness of this derivative means that, for each fixed $B$,
there is a finite constant $c_B$ with
$\|DF(B)[C]\|\le c_B\|C\|$.

\paragraph{The fixed point theorem used here.}
We use the following form of the Earle--Hamilton theorem. Let
$\mathcal X$ be a complex Banach space and let $\mathcal U$ be a
nonempty bounded convex open subset of $\mathcal X$. Suppose
$F:\mathcal U\to\mathcal U$ is holomorphic and there is a number
$\epsilon>0$ such that
\begin{equation}
\{Y\in\mathcal X:\|Y-F(X)\|<\epsilon\}\subset\mathcal U
\qquad\text{for every }X\in\mathcal U.
\end{equation}
Then $F$ has exactly one fixed point in $\mathcal U$, and its
successive iterates converge to that point in norm from every
starting point in $\mathcal U$. See \cite[Theorem~3.1]{HRS}
(p.~5 of the cited arXiv version) and the original theorem
\cite{EH}. The displayed condition means that every image point
has an entire norm ball of the same radius $\epsilon$ inside the
domain. We verify each hypothesis below for a map whose points
are themselves bounded operators.

\paragraph{1. The Banach space and the domain.}
Take $\mathcal X=\BB(\HH)$ with $\HH=L^2(\mathbb R^3)$.
This is a complex Banach space with the operator norm: a Cauchy
sequence of bounded operators in that norm converges to a bounded
operator in the same norm.
For fixed $z=E+i\tau$ with $\tau>0$, let $\Lambda_\tau>0$ be a
finite radius, to be specified in Step~5. Consider the
bounded convex open subset of the complex Banach space $\BB$
\begin{equation}\label{eq:fixed-point-domain}
\mathcal D_\tau=\{B:\|B\|<\Lambda_\tau,\quad
\im B>-(\tau/2)I\}.
\end{equation}
The strict inequality means that $\im B+(\tau/2)I$ is bounded below
by some positive multiple of $I$. The domain is nonempty because
$0\in\mathcal D_\tau$, and it is bounded by its norm restriction.
It is convex: for $B_1,B_2\in\mathcal D_\tau$ and $0\le t\le1$,
the norm inequality is preserved by $tB_1+(1-t)B_2$, and
\begin{equation}
\im\bigl(tB_1+(1-t)B_2\bigr)+(\tau/2)I
=t\bigl(\im B_1+(\tau/2)I\bigr)
 +(1-t)\bigl(\im B_2+(\tau/2)I\bigr)>0.
\end{equation}
To check openness, fix $B\in\mathcal D_\tau$ and choose
$\delta_B>0$ with $\im B+(\tau/2)I\ge\delta_B I$.
If an operator perturbation $C$ satisfies
\begin{equation}
\|C\|<\tfrac12\min\{\Lambda_\tau-\|B\|,\delta_B\},
\end{equation}
then $\|B+C\|<\Lambda_\tau$ and
\begin{equation}
\im(B+C)+(\tau/2)I
\ge(\delta_B-\|C\|)I\ge(\delta_B/2)I>0.
\end{equation}
Here we used $\|\im C\|\le\|C\|$. Hence a norm neighborhood of
each $B$ lies in $\mathcal D_\tau$, proving openness. Convexity
also gives connectedness, as required of a domain.

\paragraph{2. The resolvent is well defined.}
Fix a self-adjoint operator $H$ on $\HH=L^2(\mathbb R^3)$.
In the homogeneous problem take $H=-\Delta$; in the exterior
problem take $H=H_W$, the self-adjoint realization of
$-\Delta+W$ specified in Section~\ref{sec:notation-assumptions}. The operator $H$ remains
fixed throughout the iteration and need not be bounded.
For $B\in\mathcal D_\tau$, we now prove that $H-z-B$ has a
bounded inverse.

Fix $B\in\mathcal D_\tau$. Bounded perturbation of the self-adjoint
operator $H$ shows that $H-z-B$ is closed on $\Dom H$ and its
adjoint is $H-\overline z-B^*$ on the same domain. For $u\in\Dom H$,
\begin{equation}
-\im\langle(H-z-B)u,u\rangle
=\tau\|u\|_{\HH}^2+\langle(\im B)u,u\rangle
\ge(\tau/2)\|u\|_{\HH}^2.
\end{equation}
By Cauchy--Schwarz,
$\|(H-z-B)u\|_{\HH}\ge(\tau/2)\|u\|_{\HH}$. The same argument applied to
the adjoint, with the opposite sign of its imaginary part, gives
$\|(H-\overline z-B^*)u\|_{\HH}\ge(\tau/2)\|u\|_{\HH}$.
The first inequality gives injectivity and closed range. For the
latter assertion, if $(H-z-B)u_j$ converges, the inequality makes
$u_j$ Cauchy; closedness then places its limit in $\Dom H$ with
the required image. The adjoint inequality makes its kernel zero,
so the orthogonal complement of the range of $H-z-B$ is zero.
Thus the range is also dense, and hence equals $\HH$.

Consequently the inverse $R_B=(H-z-B)^{-1}$ exists and
\begin{equation}\label{eq:fixed-point-resolvent-estimates}
\|R_B\|\le2/\tau,\qquad
\im R_B=R_B^*(\tau I+\im B)R_B\ge0.
\end{equation}

\paragraph{3. The operator map \texorpdfstring{$B\mapsto R_B$}{B to R B} is bounded and holomorphic.}
We verify holomorphy by computing the Fr\'echet derivative and
estimating the remainder. The parameter $z$ remains fixed.
First consider the resolvent as a function of $B$.
Let $C\in\BB(\HH)$ be a perturbation, small enough
that $B+C\in\mathcal D_\tau$ and $\|C\|\|R_B\|<1$.
The resolvent identity and the norm-convergent geometric series give
\begin{equation}
R_{B+C}-R_B=R_{B+C}CR_B,\qquad
R_{B+C}=R_B(I-CR_B)^{-1}
=R_B\sum_{j=0}^{\infty}(CR_B)^j.
\end{equation}
In particular, if $\|C\|\|R_B\|\le1/2$, then
$\|R_{B+C}\|\le2\|R_B\|$ and
\begin{equation}
R_{B+C}-R_B-R_BCR_B=R_{B+C}CR_BCR_B,\qquad
\|R_{B+C}-R_B-R_BCR_B\|\le2\|R_B\|^3\|C\|^2.
\end{equation}
The map $C\mapsto R_BCR_B$ is complex-linear and bounded, since
$\|R_BCR_B\|\le\|R_B\|^2\|C\|$. The last remainder estimate
therefore proves that $B\mapsto R_B$ is holomorphic and that its
Fr\'echet derivative, applied to the direction $C$, is
\begin{equation}
D_B R_B[C]=R_BCR_B.
\end{equation}
All these statements concern operator norm. This is the
holomorphy of $R_B$ in the independent variable $B$ needed for the
fixed point argument below.

\paragraph{4. Fixed point map.}
To define the fixed point map, fix a measurable center set
$A\subset\mathbb R^3$ and use the covariance map from
\eqref{eq:covariance-map-definition}:
\begin{equation}\label{eq:construction-covariance-map}
\SSS_A[X]=\sigma^2\int_A v_aXv_a\dd a,
\qquad X\in\BB(\HH).
\end{equation}
Here $v_a$ is the multiplication operator in \eqref{eq:single-site-translate},
and the integral is understood in the weak operator sense.
The input $X$ and the output $\SSS_A[X]$ are bounded operators on
$\HH=L^2(\mathbb R^3)$. Section~\ref{sub:covariance-properties}
shows that $\SSS_A$ is bounded, complex-linear, and positivity
preserving, with $\|\SSS_A\|\le\sigma^2\|v\|_{\HH}^2$.
In the homogeneous problem $A=\mathbb R^3$ and
$\SSS_A=\SSS$; in the exterior problem $A=\Omega^c$ and
$\SSS_A=\Se$. The set $A$, and hence the covariance map
$\SSS_A$, remain fixed as $B$ varies.

Now define the fixed point map
\begin{equation}\label{eq:fixed-point-map-definition}
\mathcal F_z:\mathcal D_\tau\longrightarrow\BB(\HH),\qquad
\mathcal F_z(B):=\SSS_A[R_B]
=\SSS_A[(H-z-B)^{-1}].
\end{equation}
For each fixed $z$, this is the map denoted by $F$ in the
abstract Earle--Hamilton theorem. Its input $B$ and output
$\mathcal F_z(B)$ are bounded operators, and the Banach space
is $\BB(\HH)$ equipped with the operator norm.
The covariance map $\SSS_A$ in
\eqref{eq:construction-covariance-map} is linear in its input $X$.
To evaluate $\mathcal F_z(B)$, first form the resolvent $R_B$
and then apply $\SSS_A$; the dependence of $R_B$ on $B$ makes
$\mathcal F_z$ generally nonlinear in $B$.
In the exterior case, \eqref{eq:fixed-point-map-definition} reads
$\mathcal F_z(B)=\Se[(H_W-z-B)^{-1}]$.

Since $\SSS_A$ is bounded,
\begin{equation}\label{eq:fixed-point-map-bound}
\|\mathcal F_z(B)\|
\le\|\SSS_A\|\,\|R_B\|
\le\frac{2\sigma^2\|v\|_{\HH}^2}{\tau}.
\end{equation}
\begin{equation}
\SSS_A[X^*]=\SSS_A[X]^*,\qquad
\SSS_A[\im X]
=\frac{\SSS_A[X]-\SSS_A[X]^*}{2i}
=\im\SSS_A[X],\qquad X\in\BB(\HH).
\end{equation}
For $B\in\mathcal D_\tau$, the domain condition
\eqref{eq:fixed-point-domain} gives
$\tau I+\im B>(\tau/2)I$.
The identity \eqref{eq:fixed-point-resolvent-estimates} in Step~2
therefore shows that $\im R_B\ge0$.
Since $\SSS_A$ is positive,
$\im\mathcal F_z(B)=\SSS_A[\im R_B]\ge0$.
Together with \eqref{eq:fixed-point-map-bound}, this gives
\begin{equation}
\sup_{B\in\mathcal D_\tau}\|\mathcal F_z(B)\|
\le2\sigma^2\|v\|_{\HH}^2/\tau,
\qquad\im\mathcal F_z(B)\ge0.
\end{equation}
These estimates hold for every finite positive radius in
\eqref{eq:fixed-point-domain}.
Calling $\mathcal F_z$ a \emph{bounded map} means precisely that
its output operator norms are uniformly bounded over this domain,
as in the displayed supremum. It does not assert linearity.
Its holomorphy follows by composing the holomorphic map
$B\mapsto R_B$ with the bounded complex-linear map $\SSS_A$;
explicitly, its derivative in an operator direction $C$ is
\begin{equation}\label{eq:fixed-point-map-derivative}
D\mathcal F_z(B)[C]=\SSS_A[R_BCR_B],\qquad
\|D\mathcal F_z(B)[C]\|
\le\|\SSS_A\|\,\|R_B\|^2\|C\|.
\end{equation}
This derivative is complex-linear in $C$ because $\SSS_A$ and
multiplication by the fixed operator $R_B$ are complex-linear;
the estimate in \eqref{eq:fixed-point-map-derivative} proves boundedness.
More explicitly, applying $\SSS_A$ to the resolvent remainder
identity above gives
\begin{equation}
\mathcal F_z(B+C)-\mathcal F_z(B)-D\mathcal F_z(B)[C]
=\SSS_A[R_{B+C}CR_BCR_B].
\end{equation}
For $\|C\|\|R_B\|\le1/2$, therefore,
\begin{equation}
\|\mathcal F_z(B+C)-\mathcal F_z(B)-D\mathcal F_z(B)[C]\|
\le2\|\SSS_A\|\,\|R_B\|^3\|C\|^2.
\end{equation}
After division by $\|C\|$, the right-hand side tends to zero.
This is exactly the operator-norm limit in the definition of
holomorphy at the beginning of this subsection.
The holomorphic dependence of the resulting fixed point on the
spectral parameter $z$ is established separately in
Section~\ref{sub:holomorphic-construction}.

\paragraph{5. One uniform radius keeps the whole image inside the domain.}
Choose
\begin{equation}\label{eq:fixed-point-domain-radius}
\Lambda_\tau=1+\frac{4\sigma^2\|v\|_{\HH}^2}{\tau}.
\end{equation}
Then $2\sigma^2\|v\|_{\HH}^2/\tau<\Lambda_\tau$, so the bounds from Step~4 give
$\mathcal F_z(\mathcal D_\tau)\subset\mathcal D_\tau$.
Define
\begin{equation}
\epsilon_\tau=
\min\left\{\frac\tau4,
\frac{\Lambda_\tau-2\sigma^2\|v\|_{\HH}^2/\tau}{2}\right\}>0.
\end{equation}
It is positive because $\tau>0$ and
\eqref{eq:fixed-point-domain-radius} gives
$\Lambda_\tau-2\sigma^2\|v\|_{\HH}^2/\tau=1+2\sigma^2\|v\|_{\HH}^2/\tau>0$.
Fix any $B\in\mathcal D_\tau$ and put $Y=\mathcal F_z(B)$.
By \eqref{eq:fixed-point-map-bound}, \eqref{eq:fixed-point-domain-radius},
and the positivity conclusion of Step~4,
\begin{equation}
\|Y\|\le\frac{2\sigma^2\|v\|_{\HH}^2}{\tau}<\Lambda_\tau,
\qquad\im Y\ge0.
\end{equation}
Such an output $Y$ is called an \emph{image point}; here it is a
bounded operator in $\BB(\HH)$. A ball centered at $Y$ is measured
in operator norm. Let $C\in\BB(\HH)$ satisfy
$\|C\|<\epsilon_\tau$.
The two conditions defining $\mathcal D_\tau$ are verified separately:
\begin{equation}
\begin{aligned}
\|Y+C\|
&\le\frac{2\sigma^2\|v\|_{\HH}^2}{\tau}+\|C\|
<\frac{2\sigma^2\|v\|_{\HH}^2}{\tau}+\epsilon_\tau
\le\frac{\Lambda_\tau+2\sigma^2\|v\|_{\HH}^2/\tau}{2}
<\Lambda_\tau,\\
\im(Y+C)+\frac\tau2 I
&\ge\left(\frac\tau2-\|C\|\right)I
\ge\frac\tau4 I>0.
\end{aligned}
\end{equation}
In the second line we used $\im Y\ge0$ and
$\im C\ge-\|C\|I$. It follows that $Y+C\in\mathcal D_\tau$.
We have therefore proved
\begin{equation}
\{Y+C:\|C\|<\epsilon_\tau\}\subset\mathcal D_\tau,
\qquad Y=\mathcal F_z(B),\quad B\in\mathcal D_\tau.
\end{equation}
The same radius $\epsilon_\tau$ works for every input $B$.
This is exactly the uniform interior-image condition in the
fixed point theorem, not merely the assertion that each image
point belongs to the open domain.

\paragraph{6. Existence, uniqueness, and norm convergence.}
We now apply the theorem with
\begin{equation}
\mathcal X=\BB(\HH),\qquad
\mathcal U=\mathcal D_\tau,\qquad F=\mathcal F_z.
\end{equation}
Step~1 verified the properties of the domain. Steps~2--4 showed
that $\mathcal F_z:\mathcal D_\tau\to\BB(\HH)$ is well defined,
bounded, and holomorphic. Step~5 showed that its image lies in
$\mathcal D_\tau$ with a uniform interior margin.
Take $H=H_W$ and $A=\Omega^c$. The theorem gives a unique fixed point
$\Sigma_e(z)\in\mathcal D_\tau$ satisfying the exterior Dyson equation
\begin{equation}
\Sigma_e(z)=\mathcal F_z(\Sigma_e(z))
=\SSS_{\Omega^c}[(H_W-z-\Sigma_e(z))^{-1}].
\end{equation}
The successive approximations converge to $\Sigma_e(z)$ in operator norm
from every initial operator in $\mathcal D_\tau$. Since $0\in\mathcal D_\tau$,
this applies in particular to the iterates $\Sigma_e^{(n)}(z)$
of \eqref{eq:iteration}, starting at $\Sigma_e^{(0)}(z)=0$.

Because the fixed point belongs to the image of $\mathcal F_z$,
it has $\im\Sigma_e(z)\ge0$. Repeating the imaginary-part estimate with
this stronger sign gives
\begin{equation}\label{eq:fixed-point-solution-bounds}
\|R_e(z)\|\le\frac1\tau,
\qquad
\|\Sigma_e(z)\|=\|\SSS_{\Omega^c}[R_e(z)]\|\le\frac{\sigma^2\|v\|_{\HH}^2}{\tau}.
\end{equation}
To obtain uniqueness among all bounded solutions with nonnegative
imaginary part, let $\widetilde\Sigma$ be any such solution. The same
two estimates give $\|\widetilde\Sigma\|\le\sigma^2\|v\|_{\HH}^2/\tau<\Lambda_\tau$
and $\im\widetilde\Sigma+(\tau/2)I\ge(\tau/2)I>0$.
Thus $\widetilde\Sigma\in\mathcal D_\tau$, where the fixed point
theorem already proves uniqueness. The extension to the wider
admissible classes stated in the main theorems is proved in
Lemmas~\ref{prop:admissible-mu} and~\ref{prop:admissible-exterior}.
Lemma~\ref{prop:admissible-mu} treats the homogeneous equation;
Lemma~\ref{prop:admissible-exterior} treats the exterior
operator equation. In each case, the equation forces an
admissible solution to be bounded, so the bounded-solution
uniqueness result above applies.

If the initial operator $X$ is bounded with $\im X\ge0$ and lies outside
$\mathcal D_\tau$, then $\|(H_W-z-X)^{-1}\|\le1/\tau$. Its first iterate
$\SSS_{\Omega^c}[(H_W-z-X)^{-1}]$ therefore has nonnegative imaginary part and norm at most
$\sigma^2\|v\|_{\HH}^2/\tau$, so it belongs to $\mathcal D_\tau$. The
successive approximations from $X$ thus converge to $\Sigma_e(z)$.

\subsubsection{Holomorphic dependence on \texorpdfstring{$z$}{z} and application to the homogeneous problem}\label{sub:holomorphic-construction}
We now let $z$ vary in $\mathbb C_+$, keeping $H_W$ and the center set $\Omega^c$ fixed. Here holomorphy concerns the scalar spectral parameter
$z$, with values in $\BB(\HH)$ and derivatives taken in operator
norm. In the exterior problem we prove this property for
$\Sigma_e(z)$ and $R_e(z)$. The application to the homogeneous problem at
the end of the subsection identifies the self-energy with
multiplication by $\mu(\xi,z)$ in momentum space and gives holomorphy of
$z\mapsto\mu(\cdot,z)$ in $L^\infty(\mathbb R^3)$.
The order of the argument is to prove holomorphy of each
successive approximation, then of their limit, and finally of the
resolvent defined using that limit.

\paragraph{Holomorphy of the successive approximations by induction.}
For the iterates in \eqref{eq:iteration}, write
\begin{equation}
\Sigma_e^{(0)}(z)=0,\qquad
R_n(z)=(H_W-z-\Sigma_e^{(n)}(z))^{-1},\qquad
\Sigma_e^{(n+1)}(z)=\SSS_{\Omega^c}[R_n(z)].
\end{equation}
The imaginary-part identity from the preceding subsection and
positivity of $\SSS_{\Omega^c}$ show inductively that
$\im \Sigma_e^{(n)}(z)\ge0$. Thus each inverse exists on the fixed domain
$\Dom H_W$ and satisfies $\|R_n(z)\|\le1/\im z$.

The initial function $\Sigma_e^{(0)}(z)=0$ is holomorphic. Suppose that
$\Sigma_e^{(n)}(z)$ is holomorphic in operator norm, and fix
$z_0\in\mathbb C_+$. Define the bounded perturbation
\begin{equation}
\Delta_n(z)=(z-z_0)I+\Sigma_e^{(n)}(z)-\Sigma_e^{(n)}(z_0).
\end{equation}
It is holomorphic and tends to zero in norm as $z\to z_0$.
For $z$ sufficiently close to $z_0$, the resolvent identity gives
\begin{equation}
R_n(z)=R_n(z_0)\bigl(I-\Delta_n(z)R_n(z_0)\bigr)^{-1}
=R_n(z_0)\sum_{j=0}^{\infty}\bigl(\Delta_n(z)R_n(z_0)\bigr)^j.
\end{equation}
On a smaller neighborhood the norm of
$\Delta_n(z)R_n(z_0)$ is bounded by a fixed number less than one.
The series therefore converges locally uniformly in operator
norm, and each term is holomorphic. It follows that $R_n(z)$
is holomorphic near $z_0$. Since $z_0$ was arbitrary, it is
holomorphic on $\mathbb C_+$. The map $\SSS_{\Omega^c}$ is bounded
and complex-linear, so $\Sigma_e^{(n+1)}(z)=\SSS_{\Omega^c}[R_n(z)]$ is
holomorphic as well. This completes the induction.

\paragraph{Local uniform convergence in operator norm.}
Our goal is to prove \eqref{eq:local-uniform-iteration}.
Fix $z\in\mathbb C_+$ and put $\tau=\im z$.
The approximations defined above satisfy $\Sigma_e^{(0)}(z)=0\in\mathcal D_\tau$
and $\Sigma_e^{(n+1)}(z)=\SSS_{\Omega^c}[R_n(z)]=\mathcal F_z(\Sigma_e^{(n)}(z))$,
where $\mathcal F_z$ is the map in \eqref{eq:fixed-point-map-definition}.
By Step~5 of Section~\ref{sub:fixed-point-construction}, all these
iterates remain in $\mathcal D_\tau$. Step~6 applies the
Earle--Hamilton theorem, whose conclusion includes convergence of
the successive iterates to the unique fixed point $\Sigma_e(z)$.
Here the Banach space is $\BB(\HH)$ with its operator norm.
Consequently, for this fixed $z$,
\begin{equation}
\|\Sigma_e^{(n)}(z)-\Sigma_e(z)\|\longrightarrow0\qquad(n\to\infty).
\end{equation}
The preceding induction proves that every $\Sigma_e^{(n)}$ is holomorphic.
Moreover, $\|\SSS_{\Omega^c}\|\le\sigma^2\|v\|_{\HH}^2$ and
$\|R_n(z)\|\le1/\im z$ give
\begin{equation}\label{eq:iterate-basic-bound}
\|\Sigma_e^{(n)}(z)\|\le\frac{\sigma^2\|v\|_{\HH}^2}{\im z},
\qquad z\in\mathbb C_+,\quad n\ge0.
\end{equation}
For $n\ge1$ this follows from $\Sigma_e^{(n)}=\SSS_{\Omega^c}[R_{n-1}]$;
for $n=0$ it follows from $\Sigma_e^{(0)}=0$.
We now show directly that these facts imply convergence
uniform on every compact subset of $\mathbb C_+$.

Fix $z_0\in\mathbb C_+$ and choose a scalar radius $\rho>0$
with $2\rho<\im z_0$. For $|\zeta-z_0|\le2\rho$,
\eqref{eq:iterate-basic-bound} gives the bound
\begin{equation}
\|\Sigma_e^{(n)}(\zeta)\|
\le\frac{\sigma^2\|v\|_{\HH}^2}{\im z_0-2\rho},
\end{equation}
independently of $n$.
If $|z-z_0|\le\rho$, the circle $|\zeta-z|=\rho$ lies
in this larger closed disc. Cauchy's estimate on this circle
therefore gives
\begin{equation}
\|\partial_z\Sigma_e^{(n)}(z)\|\le
L_{z_0,\rho}:=
\frac{\sigma^2\|v\|_{\HH}^2}{\rho(\im z_0-2\rho)},
\qquad |z-z_0|\le\rho.
\end{equation}
This operator-norm estimate follows from the ordinary scalar
Cauchy estimate applied to $\langle \Sigma_e^{(n)}(z)f,g\rangle$, followed
by taking the supremum over unit vectors $f,g\in\HH$.

The smaller closed disc is convex. Integrating the derivative
along the segment $w+t(z-w)$, $0\le t\le1$, gives
\begin{equation}
\Sigma_e^{(n)}(z)-\Sigma_e^{(n)}(w)
=\int_0^1 \partial_z\Sigma_e^{(n)}(w+t(z-w))(z-w)\dd t,
\end{equation}
and hence
\begin{equation}
\|\Sigma_e^{(n)}(z)-\Sigma_e^{(n)}(w)\|\le L_{z_0,\rho}|z-w|,
\qquad |z-z_0|,|w-z_0|\le\rho.
\end{equation}
Passing to the pointwise operator-norm limit at $z$ and $w$
shows that the same bound holds for $\Sigma_e$:
\begin{equation}
\|\Sigma_e(z)-\Sigma_e(w)\|\le L_{z_0,\rho}|z-w|.
\end{equation}
This step uses only the already proved pointwise convergence;
holomorphy of $\Sigma_e$ has not yet been assumed.

Let $h>0$. Compactness of the smaller closed disc provides
finitely many points $z_1,\ldots,z_N$ in it such that every
$z$ in the disc satisfies $|z-z_j|<h$ for at least one $j$.
For such a point $z_j$, the triangle inequality gives
\begin{equation}
\begin{aligned}
\|\Sigma_e^{(n)}(z)-\Sigma_e(z)\|
&\le\|\Sigma_e^{(n)}(z)-\Sigma_e^{(n)}(z_j)\|
  +\|\Sigma_e^{(n)}(z_j)-\Sigma_e(z_j)\|+\|\Sigma_e(z_j)-\Sigma_e(z)\|\\
&\le2L_{z_0,\rho}h
  +\max_{1\le j\le N}\|\Sigma_e^{(n)}(z_j)-\Sigma_e(z_j)\|.
\end{aligned}
\end{equation}
Given $\varepsilon>0$, first choose $h$ so small that
$2L_{z_0,\rho}h<\varepsilon/2$. For this fixed finite covering,
pointwise norm convergence at the finitely many points $z_j$
then makes the maximum less than $\varepsilon/2$ for all
sufficiently large $n$. Therefore
\begin{equation}
\sup_{|z-z_0|\le\rho}\|\Sigma_e^{(n)}(z)-\Sigma_e(z)\|\longrightarrow0.
\end{equation}
Every compact set $K\subset\mathbb C_+$ is covered by finitely
many such smaller discs. We conclude that
\begin{equation}\label{eq:local-uniform-iteration}
\sup_{z\in K}\|\Sigma_e^{(n)}(z)-\Sigma_e(z)\|\longrightarrow0
\quad\text{for every compact }K\subset\mathbb C_+.
\end{equation}

\paragraph{Holomorphy of the limiting self-energy in \texorpdfstring{$z$}{z}.}
Each $\Sigma_e^{(n)}$ is holomorphic, and
\eqref{eq:local-uniform-iteration} proves locally uniform
convergence in operator norm. A locally uniform norm limit of
holomorphic Banach-space-valued functions is holomorphic; see,
for example, \cite[Theorem~2.1, p.~785]{ArendtNikolski}.
Thus $\Sigma_e(z)$ is holomorphic in operator norm throughout
$\mathbb C_+$.

\paragraph{Holomorphy of the exterior resolvent in \texorpdfstring{$z$}{z}.}
We now use the holomorphy of $\Sigma_e$ already established above.
For fixed $z\in\mathbb C_+$ and a small complex increment $h$
with $z+h\in\mathbb C_+$, the resolvent identity reads
\begin{equation}
\begin{aligned}
R_e(z+h)-R_e(z)
={}&R_e(z+h)\bigl[hI+\Sigma_e(z+h)-\Sigma_e(z)\bigr]R_e(z).
\end{aligned}
\end{equation}
By \eqref{eq:fixed-point-solution-bounds} and the norm continuity
of $\Sigma_e$, the displayed identity first gives
$R_e(z+h)\to R_e(z)$ in operator norm. Dividing by $h$ and
using norm differentiability of $\Sigma_e$ then gives
\begin{equation}
\begin{aligned}
\frac{R_e(z+h)-R_e(z)}{h}
&=R_e(z+h)
\left[I+\frac{\Sigma_e(z+h)-\Sigma_e(z)}{h}\right]R_e(z)\\
&\longrightarrow
R_e(z)\bigl(I+\Sigma_e'(z)\bigr)R_e(z)
\quad\text{in operator norm}.
\end{aligned}
\end{equation}
Thus $R_e$ is holomorphic on $\mathbb C_+$, with
\begin{equation}
R_e'(z)=R_e(z)\bigl(I+\Sigma_e'(z)\bigr)R_e(z).
\end{equation}
Here the prime denotes the complex derivative in operator norm.

\paragraph{Application to the homogeneous problem and the momentum variable.}
Take $\Omega=\varnothing$ and $W=0$. A bounded translation-invariant operator
$\Sigma_0(z)$ with nonnegative imaginary part has the form
$\Sigma_0(z)=\mathcal F^{-1}M_{\mu(\cdot,z)}\mathcal F$, where
$\mu(\cdot,z)\in L^\infty(\mathbb R^3)$ and $\im\mu(\cdot,z)\ge0$ almost
everywhere. Next, we derive the equation for $\mu$. For $f\in L^\infty(\mathbb R^3)$,
write $A_f=\mathcal F^{-1}M_f\mathcal F$, so that $\|A_f\|=\|f\|_\infty$.
For every $u\in\HH$, the homogeneous covariance map satisfies
\begin{equation}\label{eq:momentum-convolution}
\widehat{\SSS[A_f]u}(\xi)
=\frac{\sigma^2}{(2\pi)^3}\!
\left(\int_{\mathbb R^3}|\widehat v(\xi-\eta)|^2 f(\eta)\dd\eta\right)
\widehat u(\xi).
\end{equation}
The derivation from \eqref{eq:covariance-map-definition} is given in
Appendix~\ref{app:homogeneous-multiplier}.

The resolvent $(-\Delta-z-A_\mu)^{-1}$ has symbol
$(|\xi|^2-z-\mu(\xi,z))^{-1}$, whose modulus is at most $1/\tau$.
Substituting this symbol for $f$ in \eqref{eq:momentum-convolution} proves that
$\Sigma_0:=A_\mu$ solves \eqref{eq:hom-dyson-operator} if and only if $\mu$ solves \eqref{eq:mu}.

For $\Omega=\varnothing$ and $W=0$, the exterior iterates $\Sigma_e^{(n)}(z)$ specialize to the homogeneous iterates $\Sigma_0^{(n)}(z)$.
Since $\Sigma_0^{(0)}(z)=0$, \eqref{eq:momentum-convolution}
gives $\Sigma_0^{(n)}(z)=A_{\mu_n(\cdot,z)}$ by induction, where
\begin{equation}\label{eq:homogeneous-scalar-iteration}
\begin{gathered}
\mu_0(\xi,z)=0,\qquad
r_n(\xi,z)=\frac1{|\xi|^2-z-\mu_n(\xi,z)},\\
\mu_{n+1}(\xi,z)
=\frac{\sigma^2}{(2\pi)^3}\int_{\mathbb R^3}
|\widehat v(\xi-\eta)|^2r_n(\eta,z)\dd\eta.
\end{gathered}
\end{equation}
Here $\im\mu_n\ge0$ and $\|r_n(\cdot,z)\|_\infty\le1/\tau$ at
each step. For a multiplier $A_f$, the fixed-point domain
\eqref{eq:fixed-point-domain} is given by
\begin{equation}\label{eq:homogeneous-scalar-domain}
\|f\|_\infty<\Lambda_\tau,\qquad
\operatorname*{ess\,inf}_{\xi\in\mathbb R^3}
\bigl(\im f(\xi)+\tau/2\bigr)>0.
\end{equation}
Step~6 of Section~\ref{sub:fixed-point-construction} gives convergence
of $\Sigma_0^{(n)}(z)$ in operator norm to $\Sigma_0(z)$. Since
\begin{equation}\label{eq:homogeneous-iterate-norm-identity}
\|\mu_n-\mu_m\|_\infty=\|\Sigma_0^{(n)}-\Sigma_0^{(m)}\|,
\end{equation}
the scalar sequence is Cauchy
in $L^\infty(\mathbb R^3)$. Denoting its limit by $\mu(\cdot,z)$ and
using \eqref{eq:iterate-basic-bound}, we obtain
\begin{equation}\label{eq:homogeneous-scalar-convergence}
\|\mu_n(\cdot,z)-\mu(\cdot,z)\|_\infty\longrightarrow0,
\qquad
\|\mu_n(\cdot,z)\|_\infty\le\frac{\sigma^2\|v\|_{\HH}^2}{\im z}.
\end{equation}
Passing to the limit in $\Sigma_0^{(n)}=A_{\mu_n}$ gives
\begin{equation}\label{eq:homogeneous-limit-operator}
\widehat{\Sigma_0(z)u}(\xi)=\mu(\xi,z)\widehat u(\xi),
\qquad u\in\HH.
\end{equation}
Thus $\Sigma_0(z)$ is translation invariant. Its symbol has
$\im\mu\ge0$ and solves \eqref{eq:mu} by \eqref{eq:momentum-convolution}.
The map $f\mapsto A_f$ is a linear isometry onto a closed subspace
of $\BB(\HH)$. Its inverse on this subspace therefore carries the
holomorphic function $z\mapsto\Sigma_0(z)$ to the holomorphic
$L^\infty$-valued function $z\mapsto\mu(\cdot,z)$.

Finally, every bounded solution $f$ of \eqref{eq:mu} with $\im f\ge0$
satisfies $\|f\|_\infty\le\sigma^2\|v\|_{\HH}^2/\tau$ and hence
\eqref{eq:homogeneous-scalar-domain}. Its operator $A_f$ solves
\eqref{eq:hom-dyson-operator}. Uniqueness in Step~6 now gives
$A_f=\Sigma_0(z)$, so $f=\mu(\cdot,z)$ almost everywhere.

For every multi-index $\alpha$, compact support and $v\in C_c^2$ give
$\partial_\xi^\alpha\widehat v=\widehat{(-ix)^\alpha v}$.
Two integrations by parts show that
$|\partial_\xi^\alpha\widehat v(\xi)|\le C_\alpha(1+|\xi|^2)^{-1}$;
the product rule therefore gives
$|\partial_\xi^\alpha(|\widehat v|^2)(\xi)|\le
C_\alpha(1+|\xi|^2)^{-2}$, in particular
$\partial_\xi^\alpha(|\widehat v|^2)\in L^1(\mathbb R^3)$.
Differentiating the convolution therefore shows
that the solution has a $C^\infty$ representative, with
$\|\partial^\alpha\mu\|_\infty\le\frac{\sigma^2}{(2\pi)^3\tau}\|\partial^\alpha(|\widehat v|^2)\|_1$.
This preliminary estimate suffices to establish smoothness for each fixed
$\tau>0$. Lemma~\ref{lem:hom-shell-lower} in
Section~\ref{sub:hom-lower} proves the stronger high-energy estimate
\begin{equation}
\|\partial_\xi^\alpha\mu(\cdot,k^2+i\tau)\|_\infty
\le \frac{C_\alpha}{k},
\qquad 0<\tau\le1,
\end{equation}
for every fixed multi-index $\alpha$ and all $k\ge k_0$, where
$k_0$ depends only on $v,\sigma^2,s$, as stated in \eqref{eq:mu-bounds}. In particular, this stronger estimate
remains uniform as $\tau\downarrow0$.
The imaginary part of $r$ is positive everywhere;
convolution with the nonzero nonnegative $|\widehat v|^2$ gives $\im\mu>0$.

\subsection{Weighted resolvent bounds for the covariance over
\texorpdfstring{$\Omega^c$}{Omega c}}
\label{app:positive}

\begin{lemma}[Resolvent bound for two covariance modes over \texorpdfstring{$\Omega^c$}{Omega c}]
\label{lem:mode-resolvent}
Under the assumptions of Theorem~\ref{thm:general-exterior}, the two-mode
estimate \eqref{eq:K-mixed-mode} holds for every $z\in\mathbb C_+$ and for both $R=R_e(z)$ and
$R=R_e(z)^*$.
\end{lemma}

\noindent{\em Remark.}
No support condition on $W$, boundedness of $\Omega$, or high-energy
assumption is required. In particular, the bound does not depend
on the number of components of $\Omega$.

\begin{proof}[Proof of Lemma~\ref{lem:mode-resolvent}]
The proof uses the resolvent factorization in
Lemma~\ref{lem:exterior-factorization}, the covariance-mode comparison
\eqref{eq:mode-order-comparison}, and operator monotonicity of the
square root.

For $z=E+i\tau\in\mathbb C_+$, write $\Sigma_e=\Sigma_e(z)$ and set
\begin{equation}\label{eq:factorization-data}
A=H_W-E-\re\Sigma_e,\qquad
\Gamma=\tau I+\im\Sigma_e\ge\tau I.
\end{equation}
Here $\Dom A=\Dom H_W$. Section~\ref{app:exist} gives
$\Sigma_e\in\BB(\HH)$ and $\im\Sigma_e\ge0$, so $\Gamma$ is bounded
and boundedly invertible. Define
\begin{equation}\label{eq:factorization-T}
T=\Gamma^{-1/2}A\Gamma^{-1/2},\qquad
\Dom T=\{u\in\HH:\Gamma^{-1/2}u\in\Dom A\}.
\end{equation}
Lemma~\ref{lem:exterior-factorization} proves that $T$ is self-adjoint.
Its spectral calculus defines the bounded operators
\begin{equation}\label{eq:balanced-def}
D=(I+T^2)^{-1/2},\qquad
Q=\Gamma^{-1/2}D^{1/2},\qquad
U=(T+i)(I+T^2)^{-1/2}.
\end{equation}
The same lemma proves that $D$ is nonnegative and injective and that
\begin{equation}\label{eq:factorization-identities}
R_e=QUQ^*,\quad U^*U=UU^*=I,\quad \im U=D>0,\quad
Q^*\Gamma Q=D,\qquad \im R_e=QDQ^*.
\end{equation}
The factorization $R_e=QUQ^*$ in \eqref{eq:factorization-identities}
is called a \emph{balanced polar decomposition}, in analogy with the
factorization in \cite[(3.1)--(3.2)]{AEK}.
The following steps prove the two asserted resolvent bounds.

\subsubsection{The transformed covariance identity}\label{sub:saturated}
Define the transformed covariance map on $\BB(\HH)$ by
\begin{equation}\label{eq:saturated}
\Phi_e[X]=Q^*\Se[QXQ^*]Q.
\end{equation}
Here $\Se$ is the exterior covariance map defined in
\eqref{eq:exterior-covariance-map}.

The main reason for introducing $\Phi_e$ is the order inequality
$\Phi_e[D]\le D$, established in \eqref{eq:transformed-dominance}.
Combined with the covariance-mode comparison \eqref{eq:mode-order-comparison},
it yields the positive-mode bound \eqref{eq:Q-mode}.

Taking imaginary parts of the
Dyson equation $\Sigma_e=\Se[R_e]$ gives
\begin{equation}\label{eq:self-consistency-im}
\im\Sigma_e=\Se[\im R_e].
\end{equation}
The balanced polar decomposition and this use of the imaginary-part
identity follow the approach of Ajanki, Erd\H{o}s, and
Kr\"uger~\cite[Section~4.2, equations~(4.30)--(4.32)]{AjEK}
and Alt, Erd\H{o}s, and
Kr\"uger~\cite[Section~3, equations~(3.1)--(3.4)]{AEK}.
In our setting, the Hamiltonian is unbounded and $D$ need not have
a bounded inverse. Lemma~\ref{lem:exterior-factorization} justifies
the factorization, and the direct positive-mode argument in
Section~\ref{sub:sandwich-mode-norm} uses the injectivity of $D$.

Using $\im R_e=QDQ^*$ and $Q^*\Gamma Q=D$ from
\eqref{eq:factorization-identities}, together with
$\Gamma=\tau I+\im\Sigma_e$, we obtain
\begin{equation}\label{eq:transformed-self-consistency}
\Phi_e[D]=D-\tau Q^*Q.
\end{equation}
Since $\tau Q^*Q\ge0$, this gives the order inequality
\begin{equation}\label{eq:transformed-dominance}
\Phi_e[D]\le D.
\end{equation}
A detailed verification of \eqref{eq:transformed-self-consistency}
is given in Section~\ref{app:transformed-covariance-identity}.

\subsubsection{Comparison with the exterior covariance map}\label{sub:extract-mode}
Covariance modes allow us to derive the spatially weighted resolvent
bound \eqref{eq:K-mixed-mode} from the order inequality
\eqref{eq:transformed-dominance}, uniformly in $\tau>0$.
Fix a real, nonnegative $h\in L^2(\Omega^c)$ with $\|h\|_{L^2(\Omega^c)}=1$.
We prove the operator inequality
\begin{equation}\label{eq:mode-order-comparison}
m_{e,h}Xm_{e,h}\le\Se[X],\qquad X\in\BB(\HH),\quad X\ge0.
\end{equation}
Here $m_{e,h}$ is the multiplication operator defined in \eqref{eq:mode-multiplier}.

\begin{samepage}
For $X\in\BB(\HH)$ with $X\ge0$ and $f\in\HH$, the triangle and
Cauchy--Schwarz inequalities give
\begin{equation}\label{eq:mode-CS-comparison}
\begin{aligned}
&\langle m_{e,h}Xm_{e,h}f,f\rangle
=\|X^{1/2}m_{e,h}f\|_{\HH}^{\,2}
=\sigma^2\left\|\int_{\Omega^c}h(a)X^{1/2}v_af\dd a\right\|_{\HH}^{\,2}\\
&\le\sigma^2\left(\int_{\Omega^c}h(a)\|X^{1/2}v_af\|_{\HH}\dd a\right)^2
\le\sigma^2\|h\|_{L^2(\Omega^c)}^2\int_{\Omega^c}\|X^{1/2}v_af\|_{\HH}^{\,2}\dd a
=\langle\Se[X]f,f\rangle.
\end{aligned}
\end{equation}
\end{samepage}
The first inequality is the triangle inequality for the vector integral,
and the second is the scalar Cauchy--Schwarz inequality. The final
equality uses the normalization $\|h\|_{L^2(\Omega^c)}=1$.

\subsubsection{Positive-mode bound}
\label{sub:sandwich-mode-norm}
We prove the bound
\begin{equation}\label{eq:Q-mode}
0\le Q^*m_{e,h} Q\le I.
\end{equation}
Set $B=Q^*m_{e,h}Q\ge0$. Applying
\eqref{eq:mode-order-comparison} to $X=QDQ^*\ge0$, and then multiplying
on the left by $Q^*$ and on the right by $Q$, gives
\begin{equation}
BDB\le Q^*\Se[QDQ^*]Q=\Phi_e[D]\le D,
\end{equation}
where the last inequality is \eqref{eq:transformed-dominance}.
Put $K=D^{1/2}BD^{1/2}\ge0$. Then
\begin{equation}
K^2=D^{1/2}BDBD^{1/2}\le D^2.
\end{equation}
By operator monotonicity of the square root
\cite[Theorem, p.~309]{Pedersen1972},
\begin{equation}
K=(K^2)^{1/2}\le(D^2)^{1/2}=D.
\end{equation}
Consequently, $D^{1/2}(I-B)D^{1/2}\ge0$, so
$\ip{(I-B)u}u\ge0$ for every $u\in\operatorname{Ran}D^{1/2}$.
Since $D$ is injective, this range is dense in $\HH$.
By continuity, the inequality holds for every $u\in\HH$.
Thus $B\le I$, proving \eqref{eq:Q-mode}.

\subsubsection{Proof of the estimate \texorpdfstring{\eqref{eq:K-mixed-mode}}{(\ref{eq:K-mixed-mode})}}\label{sub:K-modes}
For $j=0,1$, apply \eqref{eq:Q-mode} to $h_j$. It gives
\begin{equation}
\|m_{e,h_j}^{1/2}Q\|^2=\|Q^*m_{e,h_j}Q\|\le1.
\end{equation}
Since $R_e=QUQ^*$ and $U$ is unitary,
\begin{equation}
\|m_{e,h_0}^{1/2}R_em_{e,h_1}^{1/2}\|
\le\|m_{e,h_0}^{1/2}Q\|\,\|Q^*m_{e,h_1}^{1/2}\|\le1.
\end{equation}
Replacing $U$ by $U^*$ proves the same estimate for $R_e^*$.
This proves \eqref{eq:K-mixed-mode} and completes the proof of
Lemma~\ref{lem:mode-resolvent}.
\end{proof}

\subsection{Uniform estimates for the self-energy}\label{app:coef}
For a nonempty bounded measurable set $U\subset\mathbb R^3$, define
\begin{equation}\label{eq:K-local-beta}
\beta_U(a):=\sup_{x\in U}v(x-a),\qquad \beta_{\varnothing}(a):=0,\qquad a\in\mathbb R^3.
\end{equation}

The self-energy $\Sigma_e(z)$ with
$\im\Sigma_e(z)\ge0$ constructed in Section~\ref{app:exist}
and its associated resolvent satisfy
\begin{equation}
R_e(z)=(H_W-z-\Sigma_e(z))^{-1},\qquad
\Sigma_e(z)=\sigma^2\int_{\Omega^c}v_aR_e(z)v_a\dd a.
\end{equation}

In Section~\ref{app:exist} we proved that $\Sigma_e(z)$ is bounded.
However, the estimate obtained there is not uniform as $\im z\to0$.
Our next goal is to prove a uniform estimate.

\begin{lemma}[Uniform estimates for the self-energy]\label{lem:uniform-coefficient}
Let $\Omega\subset\mathbb R^3$
be measurable and let $W\in L^\infty_{\mathrm{loc}}(\mathbb R^3)$
be real, with $W(x)\ge-C_W(1+|x|^2)$ a.e. for some finite
$C_W\ge0$. Let $\Sigma_e(z)$ be the exterior self-energy constructed in
Section~\ref{app:exist}. The estimates \eqref{eq:general-coefficient-bound} and
\eqref{eq:K-coverage} hold for every $z\in\mathbb C_+$ with the same constant
$C_v$, depending only on $v,\sigma^2,s$.
\end{lemma}

\begin{corollary}\label{cor:homogeneous-coefficient-bound}
The lemma also holds for $\Omega=\varnothing$. If $W=0$ as well, then
$\Sigma_e=\Sigma_0$ and $m_e\equiv m$, so \eqref{eq:general-coefficient-bound} gives
\begin{equation}
\|\mu(\cdot,z)\|_\infty=\|\Sigma_0(z)\|\le C_v,\qquad z\in\mathbb C_+.
\end{equation}
\end{corollary}

\begin{corollary}[Finite range of the exterior self-energy]\label{cor:exterior-finite-range}
Let $\Omega$ and $W$ satisfy the assumptions of Lemma~\ref{lem:uniform-coefficient},
and fix $z\in\mathbb C\setminus\mathbb R$.
For measurable spatial sets $U_1,U_2$,
\begin{equation}\label{eq:exterior-finite-range}
\one_{U_1}\Sigma_e(z)\one_{U_2}=0
\quad\text{if }\dist(U_1,U_2)>2s.
\end{equation}
\end{corollary}
\begin{proof}[Proof of the corollary]
Suppose that
$\dist(U_1,U_2)>2s$. Since $\supp v_a\subset B(a,s)$, this ball
cannot meet both $U_1$ and $U_2$. For each center point $a$, at least
one of the multiplication operators $\one_{U_1}v_a$ and
$v_a\one_{U_2}$ is therefore zero. The weak covariance identity gives
\begin{equation}
\one_{U_1}\Sigma_e(z)\one_{U_2}
=\sigma^2\int_{\Omega^c}(\one_{U_1}v_a)R_e(z)(v_a\one_{U_2})\dd a=0.
\end{equation}
\end{proof}

\noindent{\em Remarks.}
In Lemma~\ref{lem:uniform-coefficient}, $\Omega$ determines the integration domain $\Omega^c$
in the self-energy equation. No relation between $\Omega$ and the
support of $W$ is required, and $\Omega$ need not be bounded.
No representation of $W$ as a sum of translates of $v$ is assumed
in this lemma. For $z=E+i\tau$, the constant is independent of
$E$, $\tau$, $\Omega$, and $W$; in particular, the estimates are
uniform as $\tau\downarrow0$.

Since $v\ge0$ and $\supp v\subset B(0,s)$, one has
$0\le m_e\le m$ and $m_e(x)=m$ whenever $\dist(x,\Omega)>s$.
If $\Omega\subset B(x_0,L)$ with $x_0\in\Omega$, this equality holds in particular for
$|x-x_0|>L+s$; the constant in the lemma is also independent of $x_0$ and $L$.

\noindent{\em Remark on related results.}
There are close antecedents in the work of Alt, Erd\H{o}s,
and Kr\"uger~\cite{AEK}. In their finite von Neumann algebra
with normalized trace $\langle\cdot\rangle$, we write $\mathfrak m$ for
their Dyson solution and $S$ for their covariance map. An analogous
uniform bound for the self-energy $S[\mathfrak m]$ follows under the
flatness assumption
\begin{equation}
 c\langle X\rangle I\le S[X]\le C\langle X\rangle I,
 \qquad X\ge0.
\end{equation}
Indeed, \cite[Proposition~3.5, equation~(3.11), p.~1439]{AEK}
bounds $\|\mathfrak m(z)\|_2$, where
$\|X\|_2=\langle X^*X\rangle^{1/2}$, and
\cite[Lemma~B.2(i), p.~1523]{AEK} bounds
$\|S\|_{2\to\mathrm{op}}$. Consequently,
\begin{equation}
 \|S[\mathfrak m(z)]\|
 \le\|S\|_{2\to\mathrm{op}}\|\mathfrak m(z)\|_2\le C,
 \qquad \im z>0.
\end{equation}
This consequence of their estimates is an analogue of
\eqref{eq:general-coefficient-bound}; all page references here refer to the
published version of~\cite{AEK}.

The present lemma concerns the unbounded Schr\"odinger operator
$H_W$ on $L^2(\mathbb R^3)$ and the spatial covariance
$\SSS_{\Omega^c}$, for which the global flatness assumption above
is unavailable. Its constant is independent of the measurable
set $\Omega$ and the admissible potential $W$.
The weighted estimate \eqref{eq:K-coverage} also uses the
nonnegativity, bounded second derivatives, and finite support of $v$.
These spatial conclusions do not follow directly from the
cited statements; the necessary arguments are given below
and in Section~\ref{app:positive}.

\begin{proof}[Proof of Lemma~\ref{lem:uniform-coefficient}]
Fix $z\in\mathbb C_+$ and suppress it in the notation.
We use Lemma~\ref{lem:mode-resolvent}, valid for both $R_e$ and $R_e^*$,
and the finite support of $v$.
\paragraph{1. Proof of the estimate \texorpdfstring{\eqref{eq:general-coefficient-bound}}{(\ref{eq:general-coefficient-bound})}.}
For $\beta_U$ as in \eqref{eq:K-local-beta}, set
\begin{equation}\label{eq:K-local-AU}
A_U^2=\int_{\Omega^c}\beta_U(a)\dd a.
\end{equation}
The function $\beta_U(a)$ is measurable in $a$: by continuity of $v$,
the supremum over $x\in U$ can
be taken over a fixed countable dense subset of $U$, and is therefore
a countable supremum of continuous functions of $a$.
Our first goal is to prove the local estimate
\begin{equation}\label{eq:K-local-coef}
\|\Sigma_e\one_U\|,\ \|\Sigma_e^*\one_U\|
\le\bigl(\sigma^2\|v\|_\infty A_U^2\bigr)^{1/2}.
\end{equation}
If $A_U=0$, then
$\int_{\Omega^c}\beta_U(a)\dd a=0$. Since $\beta_U\ge0$, this implies
$\beta_U(a)=0$ for almost every $a\in \Omega^c$. For such $a$, the
inequality $0\le v(x-a)\le\beta_U(a)=0$ for $x\in U$ gives
\begin{equation}
 v_a\one_U=\one_Uv_a=0
 \quad\text{for almost every }a\in\Omega^c,
 \qquad\text{if }A_U=0.
\end{equation}
Here both products are multiplication by $v(x-a)\one_U(x)$.
For every center $a$ with $\beta_U(a)=0$, this function vanishes
on $U$ because $v(x-a)=0$, and outside $U$ because $\one_U(x)=0$.
Substituting this identity into the Dyson equation yields
\begin{equation}
 \Sigma_e\one_U
 =\sigma^2\int_{\Omega^c}v_aR_ev_a\one_U\dd a=0,
 \qquad\text{if }A_U=0.
\end{equation}
The same calculation with $R_e^*$ in place of $R_e$ gives
$\Sigma_e^*\one_U=0$ when $A_U=0$.
Thus the local estimate \eqref{eq:K-local-coef} holds immediately
when $A_U=0$.

Now suppose $A_U>0$ and set $h_U(a)=A_U^{-1}\sqrt{\beta_U(a)}$.
Then $h_U\ge0$ and $\|h_U\|_{L^2(\Omega^c)}=1$.
For the corresponding mode $m_{e,h_U}$ defined by \eqref{eq:mode-multiplier},
we apply \eqref{eq:K-mixed-mode} proven in Section~\ref{sub:K-modes}, with $h_0=h_1=h_U$.
For $R=R_e$ and $\Sigma=\Sigma_e$, the Dyson
equation $\Sigma=\sigma^2\int_{\Omega^c}v_aRv_a\dd a$ gives, for $f,g\in\HH$,
\begin{equation}\label{eq:K-local-covariance-pairing}
\langle f,\Sigma\one_Ug\rangle
=\sigma^2\int_{\Omega^c}\langle v_af,Rv_a\one_Ug\rangle\dd a.
\end{equation}
When $h_U(a)=0$, also $\beta_U(a)=0$, so $v_a\one_Ug=0$ and that
center contributes nothing to the integral. On $\{a\in \Omega^c:h_U(a)>0\}$, define
\begin{equation}
f_a=m_{e,h_U}^{-1/2}v_af,\qquad
g_a=m_{e,h_U}^{-1/2}v_a\one_Ug,\qquad
T_{h_U}=m_{e,h_U}^{1/2}Rm_{e,h_U}^{1/2}.
\end{equation}
Here $m_{e,h_U}^{-1/2}$ is multiplication by $m_{e,h_U}(x)^{-1/2}$ where
$m_{e,h_U}(x)>0$, and by zero where $m_{e,h_U}(x)=0$. With this choice, \eqref{eq:K-mixed-mode} gives $\|T_{h_U}\|\le1$.
Since $v_af=m_{e,h_U}^{1/2}f_a$ and
$v_a\one_Ug=m_{e,h_U}^{1/2}g_a$ for almost every such $a$, it follows that
\begin{equation}
|\langle v_af,Rv_a\one_Ug\rangle|
=|\langle f_a,T_{h_U}g_a\rangle|
\le\|f_a\|_{\HH}\,\|g_a\|_{\HH}.
\end{equation}
Apply the triangle inequality to the integral in
\eqref{eq:K-local-covariance-pairing}, use the preceding bound on its
integrand, and write
$\|f_a\|_{\HH}\,\|g_a\|_{\HH}=(\sqrt{h_U(a)}\,\|f_a\|_{\HH})
(\|g_a\|_{\HH}/\sqrt{h_U(a)})$. The ordinary Cauchy--Schwarz inequality
in the variable $a$ now gives
\begin{equation}\label{eq:K-weighted-cauchy-schwarz}
|\langle f,\Sigma\one_Ug\rangle|
\le \sigma^2
\left(\int_{\Omega^c\cap\{h_U>0\}}h_U(a)\|m_{e,h_U}^{-1/2}v_af\|_{\HH}^2\dd a\right)^{1/2}
\left(\int_{\Omega^c\cap\{h_U>0\}}
 \frac{\|m_{e,h_U}^{-1/2}v_a\one_Ug\|_{\HH}^2}{h_U(a)}\dd a\right)^{1/2}.
\end{equation}
This is the weighted Cauchy--Schwarz estimate used here. Denote the first integral by $I_1$.
The squared $\HH$ norm is an integral in the spatial variable $x$.
Since the inverse weight is defined to be zero on $\{m_{e,h_U}=0\}$,
and the integrand is measurable and nonnegative, we obtain
\begin{equation}
I_1=\int_{\{m_{e,h_U}>0\}}\frac{|f(x)|^2}{m_{e,h_U}(x)}
  \left(\int_{\Omega^c\cap\{h_U>0\}}
  h_U(a)v(x-a)^2\dd a\right)\dd x.
\end{equation}
For each fixed $x$, use $0\le v(x-a)\le\|v\|_\infty$ to estimate
the inner center integral:
\begin{equation}
\int_{\Omega^c\cap\{h_U>0\}}h_U(a)v(x-a)^2\dd a
\le\|v\|_\infty\int_{\Omega^c}h_U(a)v(x-a)\dd a
=\frac{\|v\|_\infty}{\sigma}m_{e,h_U}(x).
\end{equation}
Here centers with $h_U(a)=0$ contribute zero, and the last equality
is the definition of $m_{e,h_U}$ in \eqref{eq:mode-multiplier}, with $h=h_U$.
Substituting this pointwise bound into the outer spatial integral gives
\begin{equation}\label{eq:K-first-integral-bound}
I_1\le\frac{\|v\|_\infty}{\sigma}\|f\|_{\HH}^2.
\end{equation}
The cancellation of $m_{e,h_U}(x)$ occurs only where it is strictly positive; no division by
zero or positive uniform lower bound on this weight is used.
We now estimate the second integral explicitly. Denote it by $I_2$.
Recall that $v_a(x)=v(x-a)$,
$\beta_U(a)=\sup_{y\in U}v(y-a)$, and
$h_U(a)=\sqrt{\beta_U(a)}/A_U$. Hence, for $x\in U$,
\begin{equation}
0\le v_a(x)\le \beta_U(a)=A_U^2h_U(a)^2.
\end{equation}
Multiplying this inequality by the nonnegative quantity $v_a(x)$
and then dividing by $h_U(a)>0$ gives
\begin{equation}
\frac{v_a(x)^2}{h_U(a)}\le A_U^2h_U(a)v_a(x).
\end{equation}
Integrating in the center variable $a$ and using the definition
$m_{e,h_U}(x)=\sigma\int_{\Omega^c}h_U(a)v_a(x)\dd a$, we obtain
\begin{equation}\label{eq:K-inner-integral}
\int_{\Omega^c\cap\{h_U>0\}}\frac{v_a(x)^2}{h_U(a)}\dd a
\le A_U^2\int_{\Omega^c}h_U(a)v_a(x)\dd a
=\frac{A_U^2}{\sigma}m_{e,h_U}(x),\qquad x\in U.
\end{equation}
Expand the squared $\HH$ norm in $I_2$ as an integral in $x$ and
interchange the two nonnegative integrals by Tonelli's theorem.
With the inverse weight defined to be zero on $\{m_{e,h_U}=0\}$,
this yields
\begin{equation}\label{eq:K-second-integral-bound}
\begin{aligned}
I_2
&=\int_{U\cap\{m_{e,h_U}>0\}}\frac{|g(x)|^2}{m_{e,h_U}(x)}
\left(\int_{\Omega^c\cap\{h_U>0\}}\frac{v_a(x)^2}{h_U(a)}\dd a\right)\dd x\\
&\le\frac{A_U^2}{\sigma}
\int_{U\cap\{m_{e,h_U}>0\}}|g(x)|^2\dd x
\le\frac{A_U^2}{\sigma}\|g\|_{\HH}^2.
\end{aligned}
\end{equation}
The factor $m_{e,h_U}(x)$ on the right-hand side of
\eqref{eq:K-inner-integral} cancels the inverse weight $1/m_{e,h_U}(x)$
on $\{m_{e,h_U}>0\}$. No positive lower bound on $m_{e,h_U}$ is needed
for this estimate. Substituting \eqref{eq:K-first-integral-bound} and
\eqref{eq:K-second-integral-bound} into \eqref{eq:K-weighted-cauchy-schwarz},
we obtain \eqref{eq:K-local-coef}.

For a unit cube $C_p$, $A_{C_p}^2\le C_v$. The range of
$\Sigma_e\one_{C_p}$ lies in $C_p+B(0,2s)$.
These enlarged cubes have overlap at most $C(1+s)^3$, where $C$ is a
numerical constant. Thus
\begin{equation}
\left\|\sum_p\Sigma_e\one_{C_p}f\right\|_{\HH}^2
\le C(1+s)^3\sum_p\left\|\Sigma_e\one_{C_p}f\right\|_{\HH}^2.
\end{equation}
Summing the squares of the local estimates \eqref{eq:K-local-coef}
therefore proves \eqref{eq:general-coefficient-bound}; see Section~\ref{app:local-to-global}
for the detailed summation. The same proof applies to $\Sigma_e^*$.

\begin{samepage}
\paragraph{2. Proof of the estimate \texorpdfstring{\eqref{eq:K-coverage}}{(\ref{eq:K-coverage})}.}
Fix a unit cube $C_p$, and let
\begin{equation}
C_p^{+s}=\{a\in\mathbb R^3:\operatorname{dist}(a,C_p)<s\}
\end{equation}
be its open $s$-neighborhood, where $s$ is the support radius of $v$.
\end{samepage}
For $0<\delta\le1$, define
\begin{equation}
U_\delta=\{x\in\mathbb R^3:m_e(x)\le\delta\}.
\end{equation}
Recalling \eqref{eq:K-local-beta}, we easily show that
\begin{equation}\label{eq:K-center-restriction}
\int_{\Omega^c}\beta_{U_\delta\cap C_p}(a)\dd a
=\int_{\Omega^c\cap C_p^{+s}}\beta_{U_\delta\cap C_p}(a)\dd a.
\end{equation}
Indeed, if $a\notin C_p^{+s}$, then $|x-a|\ge s$ for every $x\in C_p$, so
$v(x-a)=0$ there. By \eqref{eq:K-local-beta} with $U=U_\delta\cap C_p$,
we obtain $\beta_{U_\delta\cap C_p}(a)=0$, which proves \eqref{eq:K-center-restriction}.

Next, take $0<\epsilon\le1$ and choose an
$\epsilon$-net $x_1,\ldots,x_{n_\epsilon}$ in $U_\delta\cap C_p$, with
$n_\epsilon\le C\epsilon^{-3}$. This means that the points $x_\ell$ belong
to that set and every point of the set is within distance
$\epsilon$ of at least one $x_\ell$.

Set $M_2=1+\|\mathcal D^2v\|_\infty$, where $\mathcal D^2v$ is the Hessian matrix of $v$
and $\|\mathcal D^2v\|_\infty$ is the supremum of its Euclidean operator norm.
Fix $a$. For each $x\in U_\delta\cap C_p$, choose an index
$\ell=\ell(x)$ with $|x-x_{\ell(x)}|\le\epsilon$ and put
$y=x_{\ell(x)}-a$, $h=x-x_{\ell(x)}$.
Taylor's formula in the two opposite directions gives
\begin{equation}
v(y\pm h)\le v(y)\pm\nabla v(y)\cdot h+\frac{M_2}{2}|h|^2.
\end{equation}
Adding these inequalities cancels the linear terms. Since
$v(y-h)\ge0$ and $|h|\le\epsilon$, we obtain
\begin{equation}
v(x-a)=v(y+h)\le2v(y)+M_2|h|^2
\le2v(x_{\ell(x)}-a)+M_2\epsilon^2
\le2\max_{1\le j\le n_\epsilon}v(x_j-a)+M_2\epsilon^2.
\end{equation}
The chosen index $\ell(x)$ may vary with $x$, but the last expression
is independent of $x$. Taking the supremum over
$x\in U_\delta\cap C_p$ and using $v\ge0$ gives
\begin{equation}
\beta_{U_\delta\cap C_p}(a)
\le2\max_{1\le\ell\le n_\epsilon}v(x_\ell-a)+M_2\epsilon^2
\le2\sum_{\ell=1}^{n_\epsilon}v(x_\ell-a)+M_2\epsilon^2.
\end{equation}
We integrate this upper bound only over $\Omega^c\cap C_p^{+s}$,
using \eqref{eq:K-center-restriction}. Since
$m_e(x_\ell)=\int_{\Omega^c}v(x_\ell-a)\dd a\le\delta$, we obtain
\begin{equation}
\begin{aligned}
\int_{\Omega^c}\beta_{U_\delta\cap C_p}(a)\dd a
&\le2\sum_{\ell=1}^{n_\epsilon}\int_{\Omega^c\cap C_p^{+s}}v(x_\ell-a)\dd a
+M_2\epsilon^2|\Omega^c\cap C_p^{+s}|\\
&\le2\sum_{\ell=1}^{n_\epsilon}m_e(x_\ell)+M_2(1+2s)^3\epsilon^2
\le2n_\epsilon\delta+M_2(1+2s)^3\epsilon^2.
\end{aligned}
\end{equation}
The restriction in \eqref{eq:K-center-restriction} is useful because
the constant error $M_2\epsilon^2$ is integrated over a set of bounded volume.
Recalling the definition \eqref{eq:K-local-AU} and identity
\eqref{eq:K-center-restriction}, we have therefore proved
\begin{equation}
A_{U_\delta\cap C_p}^{2}
=\int_{\Omega^c}\beta_{U_\delta\cap C_p}(a)\dd a
\le C_v(\epsilon^{-3}\delta+\epsilon^2).
\end{equation}
Choose $\epsilon=\delta^{1/5}$. Then
$A_{U_\delta\cap C_p}^{2}\le C_v\delta^{2/5}$, and applying
\eqref{eq:K-local-coef} with $U=U_\delta\cap C_p$ gives
\begin{equation}
\|\Sigma_e\one_{U_\delta\cap C_p}\|,\quad
\|\Sigma_e^*\one_{U_\delta\cap C_p}\|
\le C_v\delta^{1/5},
\end{equation}
uniformly in $p$. To sum these bounds, let $T$ be either $\Sigma_e$ or
$\Sigma_e^*$ and set $f_p=\one_{U_\delta\cap C_p}f$ for $f\in\HH$.
Since the unit cubes partition $\mathbb R^3$, $\sum_p f_p=\one_{U_\delta}f$
in $\HH$. By \eqref{eq:general-spatial-support} and its adjoint, $Tf_p$
is supported in $\{x:\dist(x,C_p)\le 2s\}$. By \eqref{eq:local-global-overlap},
each point belongs to at most $N_s$ of these enlarged cubes. Applying
Cauchy--Schwarz to the sum at each point, integrating, and then using
the local bound above gives
\begin{equation}
\begin{aligned}
\|T\one_{U_\delta}f\|_{\HH}^2
&=\Big\|\sum_p Tf_p\Big\|_{\HH}^2
\le N_s\sum_p\|Tf_p\|_{\HH}^2\\
&\le N_s C_v^2\delta^{2/5}\sum_p\|f_p\|_{\HH}^2
=N_s C_v^2\delta^{2/5}\|\one_{U_\delta}f\|_{\HH}^2.
\end{aligned}
\end{equation}
The last equality uses the disjointness of the cubes $C_p$.
Taking square roots and adding the two operator bounds gives, after
enlarging $C_v$,
\begin{equation}
\|\Sigma_e\one_{U_\delta}\|+
\|\Sigma_e^*\one_{U_\delta}\|\le C_v\delta^{1/5}.
\end{equation}
Next, we explain the argument for $\Sigma_e$. The identity
\begin{equation}
u=\one_{U_\delta}u+\one_{U_\delta^c}u
\end{equation}
separates $u$ into its restrictions to $U_\delta$ and its complement.
By linearity and the triangle inequality,
\begin{equation}
\|\Sigma_eu\|_{\HH}
\le\|\Sigma_e\one_{U_\delta}u\|_{\HH}
+\|\Sigma_e\one_{U_\delta^c}u\|_{\HH}.
\end{equation}
For the first term, the preceding bound on
$\Sigma_e\one_{U_\delta}$ gives
\begin{equation}
\|\Sigma_e\one_{U_\delta}u\|_{\HH}
\le C_v\delta^{1/5}\|u\|_{\HH}.
\end{equation}
For the second term, since $U_\delta^c=\{x:m_e(x)>\delta\}$,
the global operator bound \eqref{eq:general-coefficient-bound} gives
\begin{equation}
\begin{aligned}
\|\Sigma_e\one_{U_\delta^c}u\|_{\HH}
&\le C_v\|\one_{U_\delta^c}u\|_{\HH}
 =C_v\left(\int_{U_\delta^c}|u(x)|^2\dd x\right)^{1/2}\\
&\le C_v\delta^{-1/2}\left(\int_{U_\delta^c}m_e(x)|u(x)|^2\dd x\right)^{1/2}
 \le C_v\delta^{-1/2}\|m_e^{1/2}u\|_{\HH}.
\end{aligned}
\end{equation}
Combining the two terms proves
\begin{equation}\label{eq:K-split-input}
\|\Sigma_eu\|_{\HH}\le C_v\delta^{1/5}\|u\|_{\HH}
+C_v\delta^{-1/2}\|m_e^{1/2}u\|_{\HH},
\qquad 0<\delta\le1.
\end{equation}

If $u=0$, the desired estimate is immediate. For $u\ne0$, define
\begin{equation}
t=\frac{\|m_e^{1/2}u\|_{\HH}}{\|u\|_{\HH}}.
\end{equation}
Then \eqref{eq:K-split-input} becomes
\begin{equation}
\|\Sigma_eu\|_{\HH}\le C_v\|u\|_{\HH}
\bigl(\delta^{1/5}+t\delta^{-1/2}\bigr).
\end{equation}
First consider the case $0<t\le1$. Choose $\delta=t^{10/7}$.
This choice satisfies $0<\delta\le1$. Both terms then equal
$t^{2/7}$, and hence
\begin{equation}
\|\Sigma_eu\|_{\HH}\le2C_v\|u\|_{\HH}t^{2/7}
=2C_v\|u\|_{\HH}^{5/7}\|m_e^{1/2}u\|_{\HH}^{2/7}.
\end{equation}
Absorbing the factor $2$ into the constant gives
\begin{equation}\label{eq:K-interpolation-selfenergy}
\|\Sigma_eu\|_{\HH}\le C_v\|u\|_{\HH}^{5/7}\|m_e^{1/2}u\|_{\HH}^{2/7}.
\end{equation}
If $t=0$, then $\|m_e^{1/2}u\|_{\HH}=0$, so
\eqref{eq:K-split-input} gives
$\|\Sigma_eu\|_{\HH}\le C_v\delta^{1/5}\|u\|_{\HH}$ for every
$0<\delta\le1$. Letting $\delta\downarrow0$ shows that
$\Sigma_eu=0$, which proves \eqref{eq:K-interpolation-selfenergy}
in this case as well. If $t>1$, use the global bound instead:
\begin{equation}
\|\Sigma_eu\|_{\HH}\le C_v\|u\|_{\HH}
\le C_v\|u\|_{\HH}t^{2/7}
=C_v\|u\|_{\HH}^{5/7}\|m_e^{1/2}u\|_{\HH}^{2/7}.
\end{equation}
Thus \eqref{eq:K-interpolation-selfenergy} holds for every $u\in\HH$.

The analogous inequality holds for $\Sigma_e^*$:
\begin{equation}
\|\Sigma_e^*u\|_{\HH}\le C_v\|u\|_{\HH}^{5/7}\|m_e^{1/2}u\|_{\HH}^{2/7}.
\end{equation}
Indeed, the same decomposition and choice of $\delta$ apply, using
$\|\Sigma_e^*\|\le C_v$ from \eqref{eq:general-coefficient-bound} and the already proved
bound $\|\Sigma_e^*\one_{U_\delta}\|\le C_v\delta^{1/5}$.
Adding the two inequalities and enlarging $C_v$ if necessary proves
\eqref{eq:K-coverage}.
\end{proof}

\subsection{High-energy properties of the homogeneous coefficient}\label{app:hom}
Here we consider the homogeneous case $\Omega=\varnothing$, $W=0$.
In Section~\ref{sub:holomorphic-construction}, the homogeneous Dyson
equation was solved and its solution $\mu(\xi,z)$ was shown to be smooth
in the momentum variable $\xi$ for each fixed $z\in\mathbb C_+$.
The uniform bound for $\mu$ was established in
Lemma~\ref{lem:uniform-coefficient} of Section~\ref{app:coef}.
We now strengthen these results by proving bounds on all momentum
derivatives that are uniform as $\im z=\tau\downarrow0$,
for $0<\tau\le1$.

\subsubsection{High-energy estimates and the radial zero}\label{sub:hom-lower}
\begin{lemma}[High-energy estimates for the homogeneous coefficient]\label{lem:hom-shell-lower}
Under the standing assumptions on $v$ and $\sigma^2$, let
$\mu(\xi)=\mu(\xi,k^2+i\tau)$ be the homogeneous solution constructed
in Section~\ref{app:exist}, and $r(\xi)$ be given by \eqref{eq:hom-resolvent-multiplier}.
There exist constants $k_0,h,c,C,C_*>0$, depending only on $v$, $\sigma^2$,
and $s$, such that for every $k\ge k_0$ and $0<\tau\le1$,
\begin{align}
\im\mu(\xi)
&=\frac{\sigma^2}{16\pi^2k}
\int_{|\eta|=k}|\widehat v(\xi-\eta)|^2\dd S(\eta)
+O(k^{-2}),\label{eq:hom-im-asymptotic}\\
\im\mu(\xi)&\ge\frac{c}{k}
\quad\text{whenever }||\xi|-k|\le4h,\label{eq:hom-im-lower}\\
\|r\|_\infty&\le Ck.\label{eq:hom-resolvent-bound}
\end{align}
Formula \eqref{eq:hom-im-asymptotic} holds for every $\xi\in\mathbb R^3$,
with a remainder uniform in $\xi$ and $0<\tau\le1$.
Furthermore, for each $\nu\in\mathbb S^2$, the equation
\begin{equation}
\rho^2-k^2-\re\mu(\rho\nu)=0,\qquad \rho\ge0,
\end{equation}
has exactly one solution $\rho_*(\nu)>0$. It satisfies
\begin{equation}
|\rho_*(\nu)-k|\le C_*/k,\qquad
\tau+\im\mu(\rho_*(\nu)\nu)\ge c/k,
\end{equation}
and the function $\nu\mapsto\rho_*(\nu)$ belongs to
$C^\infty(\mathbb S^2)$ for each fixed $k,\tau$ in the stated range.
All constants are uniform in
$\tau$ and in the direction $\nu$ in the stated range.

Moreover, for every multi-index $\alpha\in\mathbb N_0^3$ there is a
constant $C_\alpha>0$, depending only on $\alpha$, $v$, $\sigma^2$, and $s$,
such that
\begin{equation}\label{eq:hom-high-energy-derivatives}
\|\partial_\xi^\alpha\mu(\cdot,k^2+i\tau)\|_\infty
\le\frac{C_\alpha}{k},
\qquad k\ge k_0,\quad 0<\tau\le1.
\end{equation}
The threshold $k_0$ is independent of $\alpha$, and the constants
$C_\alpha$ are independent of $k$ and $\tau$.
\end{lemma}
\begin{proof}
\begin{samepage}
We divide the proof into seven steps.

\paragraph{Step 1. Complex extension and uniform bounds.}
By Lemma~\ref{lem:uniform-coefficient}, $\|\mu\|_\infty\le C_0$
uniformly in $k,\tau$. Write $\check\mu=\mathcal F^{-1}_{\xi\to x}\mu$.
By \eqref{eq:mu-spatial-support} of Theorem~\ref{thm:mu},
$\supp\check\mu\subseteq\overline{B(0,2s)}$.
Choose $\chi\in C_c^\infty(\mathbb R^3)$, depending only on $s$,
equal to one on a neighborhood of this ball, and put
$\psi=(2\pi)^{-3}\widehat\chi$.
Since $\chi\check\mu=\check\mu$, Fourier transformation gives
$\mu=\psi*\mu$. Thus
\begin{equation}
\mu(\zeta)=\int_{\mathbb R^3}\psi(\zeta-\eta)\mu(\eta)\dd\eta,
\qquad \zeta\in\mathbb C^3,
\end{equation}
\end{samepage}
defines an entire extension of $\mu$. Indeed, $\psi$ and its derivatives
decrease rapidly in the real variables, uniformly on each fixed complex
strip. The uniform $L^1$ bounds for $\psi$ and its first derivatives on
$|\im\zeta|\le1$ therefore give
\begin{equation}\label{eq:hom-complex-strip}
|\mu(\zeta)|+|\nabla_\zeta\mu(\zeta)|\le M
\qquad (|\im\zeta|\le1),
\end{equation}
where $M$ depends only on $v,\sigma^2,s$ and is independent of $k,\tau$.

\paragraph{Step 2. The radial zero.}
Set
\begin{equation}
a(\eta)=|\eta|^2-k^2-\re\mu(\eta),\qquad
\gamma(\eta)=\tau+\im\mu(\eta).
\end{equation}
We will show that, for each direction $\nu\in\mathbb S^2$,
the function $\rho\mapsto a(\rho\nu)$ has a unique positive zero,
which we denote by $\rho_*(\nu)$, satisfying
\begin{equation}
|\rho_*(\nu)-k|\le C_*/k.
\end{equation}
This means:
\begin{equation}
\rho_*(\nu)^2-k^2-\re\mu(\rho_*(\nu)\nu)=0.
\end{equation}
The constant $C_*>0$ is independent of $k$, $\nu$, and $\tau$
in the range under consideration. The bound locates the zero within
distance $C_*/k$ of $k$; uniqueness holds on the entire half-line
$\rho\ge0$, as shown below.

To prove existence and uniqueness, use the bounds
$\|\mu\|_\infty\le C_0$ and $\|\nabla\mu\|_\infty\le M$
and choose $C_*=C_0+1>C_0$.
We may increase $k_0$ as needed, depending only on $v,\sigma^2,s$.
Taking $k_0>C_*$ ensures $C_*^2/k^2<1$ for $k\ge k_0$, so
\begin{equation}
a((k+C_*/k)\nu)\ge2C_*+C_*^2/k^2-C_0>0,
\qquad
a((k-C_*/k)\nu)\le-2C_*+C_*^2/k^2+C_0<0.
\end{equation}
Continuity therefore gives a zero between these two radii.
Moreover,
\begin{equation}\label{eq:hom-radial-derivative}
\partial_\rho a(\rho\nu)
=2\rho-\nu\cdot\nabla\re\mu(\rho\nu),\qquad
k\le\partial_\rho a(\rho\nu)\le3k
\end{equation}
on $|\rho-k|\le H$, for any fixed $H>0$, provided $k\ge2H+M$.
Here the lower bound also depends on the chosen width $H$.
The widths used below are chosen in terms of $v,\sigma^2,s$, so this
condition can be included in $k\ge k_0$.
Strict monotonicity makes the zero unique in that neighborhood.
To exclude zeros elsewhere, observe that any zero with $\rho\ge0$
must satisfy
\begin{equation}
|\rho-k|=\frac{|\rho^2-k^2|}{\rho+k}
=\frac{|\re\mu(\rho\nu)|}{\rho+k}
\le C_0/k.
\end{equation}
Hence every nonnegative zero lies in the interval where strict
monotonicity holds. The zero is therefore unique on $[0,\infty)$
and is positive for $k\ge k_0$.
The implicit function theorem makes $\rho_*(\nu)$ a continuous
(in fact $C^\infty$) function of direction.

\paragraph{Step 3. The free comparison and the Sokhotski--Plemelj formula.}
We prove the imaginary-part lower bound by replacing
$\mu(\eta)+i\tau$ in the denominator of \eqref{eq:mu} by $i0$
and estimating the error after integration. Put
$c_\sigma=\sigma^2/(2\pi)^3$ and
$F(\zeta)=\widehat v(\zeta)\widehat v(-\zeta)$.
Since $v$ is real, $F(\xi)=|\widehat v(\xi)|^2$ for real $\xi$.
Define the free comparison by
\begin{equation}\label{eq:hom-free-comparison}
\mu_{\mathrm{free}}(\xi,k)
:=c_\sigma\lim_{\varepsilon\downarrow0}
\int_{\mathbb R^3}
\frac{F(\xi-\eta)}{|\eta|^2-k^2-i\varepsilon}\dd\eta.
\end{equation}
We will justify the limit and prove
\begin{equation}\label{eq:hom-free-error}
\sup_{\xi\in\mathbb R^3}
|\mu(\xi,k^2+i\tau)-\mu_{\mathrm{free}}(\xi,k)|
\le \frac{C}{k^2},\qquad 0<\tau\le1.
\end{equation}

We first justify the limit in \eqref{eq:hom-free-comparison}.
The function $F$ is entire in complex momentum. Since $v\in C_c^2(\mathbb R^3)$, we have
\begin{equation}\label{eq:hom-complex-numerator}
|F(\xi+it)|\le C(1+|\xi|^2)^{-2},
\qquad \xi,t\in\mathbb R^3,\quad |t|\le1.
\end{equation}

By Sokhotski--Plemelj formula:
\begin{equation}
\mu_{\mathrm{free}}(\xi,k)
=c_\sigma\operatorname{p.v.}\!\int_{\mathbb R^3}
\frac{F(\xi-\eta)}{|\eta|^2-k^2}\dd\eta
+\frac{i\pi c_\sigma}{2k}
\int_{|\eta|=k}|\widehat v(\xi-\eta)|^2\dd S(\eta).
\end{equation}
Here $\dd S$ denotes surface measure. The principal-value term is real.

\paragraph{Step 4. Contour deformation and the free integral.}
We now prove \eqref{eq:hom-free-error}.
For $\nu\in\mathbb S^2$, set
\begin{equation}
D_\nu(w)=w^2-k^2-i\tau-\mu(w\nu),\qquad D_0(w)=w^2-k^2.
\end{equation}
We keep the spectral parameter $z=k^2+i\tau\in\mathbb C_+$ fixed and deform
only the radial momentum $w$, using the entire extension $w\mapsto\mu(w\nu,z)$
from Step~1. For $\varepsilon>0$, the zero of $D_0(w)-i\varepsilon$ with
positive real part lies above the real axis, so we deform the positive radial
half-line downward. Consider the closed half-strip in the complex radial momentum plane
\begin{equation}
\{\,w=t-i\beta:\ t\ge0,\ 0\le\beta\le1\,\}.
\end{equation}
We evaluate $\mu((t-i\beta)\nu,z)$ while keeping $\im z=\tau>0$.
Let $\mathcal C$ be the contour running from $0$ to $-i$ and then
along $w=t-i$, $0\le t<\infty$.
Recall that $M$ is the constant in
\eqref{eq:hom-complex-strip}: it bounds both $|\mu|$ and
$|\nabla_\zeta\mu|$ throughout the unit complex strip and is
independent of $k,\tau,t,\beta,\nu$.
The gradient bound in \eqref{eq:hom-complex-strip} gives
\begin{equation}
|\mu((t-i\beta)\nu)-\mu(t\nu)|\le M\beta,
\qquad
\im\mu((t-i\beta)\nu)\ge\im\mu(t\nu)-M\beta.
\end{equation}
Using $\re(w^2)=t^2-\beta^2$, $\im(w^2)=-2t\beta$, and
$\im\mu(t\nu)\ge0$, we obtain
\begin{equation}
\begin{aligned}
\re D_\nu(t-i\beta)&\le-\tfrac34 k^2+M<0
&& (0\le t\le k/2),\\
\im D_\nu(t-i\beta)&\le-(2t-M)\beta-\tau-\im\mu(t\nu)
&& (t\ge0).
\end{aligned}
\end{equation}
The right-hand side of the second line is strictly negative for
$t\ge M/2$, since $\tau>0$ and $\im\mu(t\nu)\ge0$.
For $k>M$, this includes $t\ge k/2$; the first line covers
$0\le t\le k/2$. Thus $D_\nu$ has no zeros throughout the
closed half-strip defined above, for sufficiently large $k$.
These sign estimates hold for every $\tau>0$.
The free denominator $D_0(w)-i\varepsilon$, $\varepsilon>0$,
also has no zeros there, since its imaginary part is
$-2t\beta-\varepsilon<0$.
Cauchy's theorem deforms both radial integrals to $\mathcal C$;
the closing segments at infinity contribute zero by
\eqref{eq:hom-complex-numerator}.
On the horizontal part $w=t-i$,
\begin{equation}
|D_0(w)-i\varepsilon|\ge|D_0(w)|
=|t-i-k|\,|t-i+k|\ge k,
\end{equation}
and on the segment $w=-i\beta$, $0\le\beta\le1$, its modulus is
at least $k^2$.
For $0<\tau\le1$, \eqref{eq:hom-complex-strip} gives
$|\mu(w\nu)+i\tau|\le M+1$ on $\mathcal C$.
Hence, for $k\ge2(M+1)$, on the horizontal part,
\begin{equation}
|D_\nu(t-i)|\ge|D_0(t-i)|-(M+1)
\ge\tfrac12|D_0(t-i)|\ge k/2.
\end{equation}
On the vertical segment,
\begin{equation}
|D_\nu(-i\beta)|\ge-\re D_\nu(-i\beta)
\ge k^2-M\ge k^2/2,\qquad 0\le\beta\le1.
\end{equation}
We next justify the following limiting contour representation:
\begin{equation}
\mu_{\mathrm{free}}(\xi,k)
=c_\sigma\lim_{\varepsilon\downarrow0}
\int_{\mathbb S^2}\int_{\mathcal C}
\frac{w^2F(\xi-w\nu)}{D_0(w)-i\varepsilon}\dd w\dd S(\nu)
=c_\sigma\int_{\mathbb S^2}\int_{\mathcal C}
\frac{w^2F(\xi-w\nu)}{D_0(w)}\dd w\dd S(\nu).
\end{equation}
To estimate the full complex integral, consider first the horizontal
part $w=t-i$. Since $|D_0(w)-i\varepsilon|\ge k$ for
$\varepsilon\ge0$, its absolute contribution is bounded by $c_\sigma/k$
times the numerator integral below. By \eqref{eq:hom-complex-numerator},
splitting at $t=1$ and using spherical coordinates for $t\ge1$, we obtain
\begin{equation}
\int_{\mathbb S^2}\int_0^\infty
(1+t^2)|F(\xi-(t-i)\nu)|\dd t\dd S(\nu)
\le C+C\int_{\mathbb R^3}
\frac{\dd\eta}{(1+|\xi-\eta|^2)^2}\le C.
\end{equation}
Here $1+t^2$ estimates the radial Jacobian on the shifted contour.
The horizontal contribution is therefore at most $C/k$.
On the vertical part $w=-i\beta$, $0\le\beta\le1$, we have
$|w^2|=\beta^2$,
$|D_0(w)-i\varepsilon|\ge|D_0(w)|=k^2+\beta^2\ge k^2$, and
$|F(\xi+i\beta\nu)|\le C$ by \eqref{eq:hom-complex-numerator}.
These bounds provide an integrable majorant independent of
$\varepsilon$ and justify the passage to the limit displayed above
by dominated convergence. They also give, uniformly in $\xi$,
\begin{equation}
|\re\mu_{\mathrm{free}}(\xi,k)|\le|\mu_{\mathrm{free}}(\xi,k)|
\le\frac{C}{k}+\frac{C}{k^2}\int_0^1\beta^2\dd\beta
\le\frac{C'}{k}.
\end{equation}
\paragraph{Step 5. The comparison estimate.}
Subtracting the two deformed integrals gives
\begin{equation}\label{eq:hom-contour-error}
\mu(\xi)-\mu_{\mathrm{free}}(\xi,k)
=c_\sigma\int_{\mathbb S^2}\int_{\mathcal C}
w^2F(\xi-w\nu)
\frac{\mu(w\nu)+i\tau}{D_\nu(w)D_0(w)}
\dd w\dd S(\nu).
\end{equation}

On the horizontal part $w=t-i$, the bounds from Step~4 give
\begin{equation}
\left|\frac{\mu(w\nu)+i\tau}{D_\nu(w)D_0(w)}\right|
\le\frac{C}{k^2}.
\end{equation}
By the uniform numerator bound established above, the horizontal
part of \eqref{eq:hom-contour-error} is $O(k^{-2})$,
uniformly in $\xi,\tau$.
On the segment $w=-i\beta$, $0\le\beta\le1$, both denominators have
modulus at least $ck^2$, whereas $|w|$, $|F(\xi-w\nu)|$, and
$|\mu(w\nu)+i\tau|$ are bounded uniformly. Its contribution is
$O(k^{-4})$. This proves \eqref{eq:hom-free-error}.

\paragraph{Step 6. Proof of \texorpdfstring{\eqref{eq:hom-im-asymptotic}--\eqref{eq:hom-resolvent-bound}}{(\ref{eq:hom-im-asymptotic})--(\ref{eq:hom-resolvent-bound})}.}
Taking imaginary parts in the Sokhotski--Plemelj formula for
$\mu_{\mathrm{free}}$ and using \eqref{eq:hom-free-error}, we obtain
\begin{equation}\label{eq:hom-free-shell}
\im\mu(\xi,k^2+i\tau)
=\frac{\pi c_\sigma}{2k}
\int_{|\eta|=k}|\widehat v(\xi-\eta)|^2\dd S(\eta)
+O(k^{-2}),
\end{equation}
uniformly in $\xi$ and $0<\tau\le1$.
This proves the asymptotic formula \eqref{eq:hom-im-asymptotic}.

Next, we prove \eqref{eq:hom-im-lower}. Since $\widehat v(0)>0$ and
$\widehat v$ is continuous, choose $0<h<1$ so that
$|\widehat v(\zeta)|\ge\widehat v(0)/2$ for $|\zeta|\le5h$.
Suppose $||\xi|-k|\le4h$ and $k>4h$, and set $\eta_0=k\xi/|\xi|$.
On the cap $|\eta|=k$, $|\eta-\eta_0|\le h$, we have
$|\xi-\eta|\le|\xi-\eta_0|+|\eta_0-\eta|\le5h$.
The cap has height $h^2/(2k)$ and hence area $\pi h^2$. It follows that
\begin{equation}
\int_{|\eta|=k}|\widehat v(\xi-\eta)|^2\dd S(\eta)
\ge \frac{\pi h^2}{4}|\widehat v(0)|^2.
\end{equation}
By \eqref{eq:hom-im-asymptotic},
$\im\mu(\xi)\ge\sigma^2 h^2|\widehat v(0)|^2/(64\pi k)-C/k^2$.
After increasing $k_0$, this gives \eqref{eq:hom-im-lower} with
$c=\sigma^2 h^2|\widehat v(0)|^2/(128\pi)$.

Let us prove \eqref{eq:hom-resolvent-bound}.
Taking $k_0$ also large enough that $C_*/k\le4h$ and $C_0\le2hk$
for $k\ge k_0$, we have $|\rho_*(\nu)-k|\le4h$.
Thus \eqref{eq:hom-im-lower} applies at $\xi=\rho_*(\nu)\nu$ and gives
\begin{equation}
\gamma(\rho_*(\nu)\nu)
=\tau+\im\mu(\rho_*(\nu)\nu)\ge c/k.
\end{equation}
This proves the asserted lower bound at the radial zero.
For the global resolvent estimate, recall
$a(\xi)=|\xi|^2-k^2-\re\mu(\xi)$ and $\gamma(\xi)=\tau+\im\mu(\xi)$, so
\begin{equation}
|r(\xi)|=\frac{1}{\sqrt{a(\xi)^2+\gamma(\xi)^2}}.
\end{equation}
If $||\xi|-k|\le4h$, then \eqref{eq:hom-im-lower} gives
$\gamma(\xi)\ge c/k$, and hence
\begin{equation}
|r(\xi)|\le\frac{1}{\gamma(\xi)}\le\frac{k}{c}.
\end{equation}
If $||\xi|-k|>4h$, the uniform bound $|\re\mu(\xi)|\le C_0$ gives
\begin{equation}
|a(\xi)|\ge\bigl||\xi|-k\bigr|(|\xi|+k)-C_0
\ge4hk-C_0\ge2hk.
\end{equation}
Therefore $|r(\xi)|\le1/|a(\xi)|\le1/(2hk)$ in this second region.
The two regions cover $\mathbb R^3$, and together these estimates
prove \eqref{eq:hom-resolvent-bound}.

\paragraph{Step 7. Upper bounds and momentum derivatives.}
Step~4 gives
$\|\mu_{\mathrm{free}}(\cdot,k)\|_\infty\le C/k$.
Combining this with \eqref{eq:hom-free-error}, we obtain
\begin{equation}
\|\mu(\cdot,k^2+i\tau)\|_\infty
\le\|\mu_{\mathrm{free}}(\cdot,k)\|_\infty+\frac{C}{k^2}
\le\frac{C'}{k},
\end{equation}
uniformly for $k\ge k_0$ and $0<\tau\le1$.

For the cutoff $\psi$ fixed in Step~1, we have
$\mu=\psi*\mu$ and $\partial^\alpha\psi\in L^1(\mathbb R^3)$ for every
multi-index $\alpha$. Hence
\begin{equation}
\partial_\xi^\alpha\mu=(\partial^\alpha\psi)*\mu,
\qquad
\|\partial_\xi^\alpha\mu\|_\infty
\le\|\partial^\alpha\psi\|_1\|\mu\|_\infty
\le\frac{C_\alpha}{k}.
\end{equation}
The cutoff $\psi$ is fixed independently of $k$ and $\tau$, so no
additional restriction on $k$ is needed for higher derivatives.
This proves \eqref{eq:hom-high-energy-derivatives}, and hence
\eqref{eq:mu-bounds}, with uniform constants.
\end{proof}

\subsubsection{The square-integrability estimate}
\begin{lemma}[Square-integrability of the homogeneous resolvent multiplier]
\label{lem:hom-resolvent-l2}
Under the assumptions of Lemma~\ref{lem:hom-shell-lower}, let $r$ be the
homogeneous resolvent multiplier defined there. There exist constants
$k_0,C>0$, depending only on $v,\sigma^2,s$, such that
\begin{equation}\label{eq:hom-resolvent-l2}
\|r(\cdot,k^2+i\tau)\|_{L^2(\mathbb R^3)}^2\le Ck^2,
\qquad k\ge k_0,\quad 0<\tau\le1.
\end{equation}
\end{lemma}

\begin{proof}
For each $\nu\in\mathbb S^2$, let $\rho_*=\rho_*(\nu)$ be the zero
from Lemma~\ref{lem:hom-shell-lower}, and recall
$a(\rho\nu)=\rho^2-k^2-\re\mu(\rho\nu)$ and
$\gamma(\rho\nu)=\tau+\im\mu(\rho\nu)$.
If $|\rho-\rho_*|\le h$, then $|\rho-k|\le2h$ for sufficiently
large $k$, since $|\rho_*-k|\le C_*/k$.
Since $a(\rho_*\nu)=0$, the radial derivative bound
\eqref{eq:hom-radial-derivative} and the lower bound
\eqref{eq:hom-im-lower} give
\begin{equation}
|a(\rho\nu)|\ge k|\rho-\rho_*|,
\qquad \gamma(\rho\nu)\ge c/k.
\end{equation}
Consequently,
\begin{equation}
|r(\rho\nu)|^2
=\frac{1}{a(\rho\nu)^2+\gamma(\rho\nu)^2}
\le\frac{C}{k^2|\rho-\rho_*|^2+k^{-2}}.
\end{equation}
Multiplying by $\rho^2\asymp k^2$ and integrating over radius
($|\rho-\rho_*|\le h$) and direction gives at most $Ck^2$.

For the remaining part of $k/2\le\rho\le2k$, use the derivative
identity in \eqref{eq:hom-radial-derivative} and
Lemma~\ref{lem:hom-shell-lower} to obtain
\begin{equation}
\partial_\rho a(\rho\nu)
=2\rho-\nu\cdot\nabla\re\mu(\rho\nu)\ge k/2
\end{equation}
throughout this interval for sufficiently large $k$.
Since $a(\rho_*\nu)=0$, it follows that
$|r|\le C/(k|\rho-\rho_*|)$. On $|\rho-\rho_*|>h$,
its squared integral with the radial weight $\rho^2$ is bounded by
a constant. Finally, for $\rho\notin[k/2,2k]$, the uniform bound
$\|\mu\|_\infty\le C/k$ from Lemma~\ref{lem:hom-shell-lower} gives
$|a(\rho\nu)|\ge c(\rho^2+k^2)$, and hence
$|r|\le C/(\rho^2+k^2)$, whose squared integral is $O(k^{-1})$.
This proves \eqref{eq:hom-resolvent-l2}.
\end{proof}

\subsection{Technical lemmas}\label{sub:technical-lemmas}
\subsubsection{The class of admissible solutions and automatic boundedness}
\label{app:admissible}

Fix $z=E+i\tau\in\mathbb C_+$, with $\tau>0$, and recall
$\HH=L^2(\mathbb R^3)$.
The following two lemmas extend the bounded-solution uniqueness from
Section~\ref{app:exist} to the admissible classes in the main theorems.
They use only the covariance bound and positivity of the imaginary part;
no high-energy assumption is required.

\paragraph{Measurable solutions of the homogeneous equation}

\begin{lemma}[Automatic boundedness of the homogeneous coefficient]
\label{prop:admissible-mu}
Let $\widetilde\mu:\mathbb R^3\to\mathbb C$ be measurable and
finite almost everywhere, with $\im\widetilde\mu\ge0$ almost
everywhere. Suppose that
\begin{equation}\label{eq:admissible-mu-dyson}
\widetilde\mu(\xi)
=\frac{\sigma^2}{(2\pi)^3}\int_{\mathbb R^3}
\frac{|\widehat v(\xi-\eta)|^2}{|\eta|^2-z-\widetilde\mu(\eta)}\dd\eta
\qquad\text{for almost every }\xi.
\end{equation}
Then
\begin{equation}\label{eq:admissible-mu-bound}
\|\widetilde\mu\|_\infty
\le\frac{\sigma^2\|v\|_{\HH}^2}{\tau}.
\end{equation}
Consequently $\widetilde\mu=\mu(\cdot,z)$ almost everywhere,
where $\mu$ is the solution of Theorem~\ref{thm:mu}.
No boundedness, decay, or integrability assumption on
$\widetilde\mu$ is needed in advance.
\end{lemma}

\begin{proof}
The imaginary part of the denominator is
$-\tau-\im\widetilde\mu(\eta)$. Hence
\begin{equation}
\bigl||\eta|^2-z-\widetilde\mu(\eta)\bigr|
\ge\tau+\im\widetilde\mu(\eta)\ge\tau
\quad\text{for almost every }\eta.
\end{equation}
Since $|\widehat v|^2\in L^1(\mathbb R^3)$, the integral on the right of
\eqref{eq:admissible-mu-dyson} is absolutely convergent and has
absolute value at most
\begin{equation}
\frac{\sigma^2}{(2\pi)^3\tau}\int_{\mathbb R^3}|\widehat v(\xi-\eta)|^2\dd\eta
=\frac{\sigma^2\|v\|_{\HH}^2}{\tau}.
\end{equation}
This proves \eqref{eq:admissible-mu-bound}. The bounded-solution
uniqueness result in Section~\ref{app:exist} now applies.
The almost-everywhere solution therefore has the smooth
representative constructed in Section~\ref{sub:holomorphic-construction}.
\end{proof}

\paragraph{Initially unbounded operator coefficients}

Let $\Omega\subset\mathbb R^3$ be measurable and let $H_W$ be the
self-adjoint operator specified in Section~\ref{sec:notation-assumptions}.
For an initially unbounded operator, the domain of the operator
and the meaning of its inverse must be specified. The following
lemma gives one such class without presupposing boundedness
of the coefficient or of the inverse.

\begin{lemma}[Automatic boundedness of an admissible operator solution]
\label{prop:admissible-exterior}
Let $\widetilde\Sigma$ be a closed, densely defined operator on
$\HH$ satisfying
\begin{equation}\label{eq:admissible-sign}
\im\langle\widetilde\Sigma u,u\rangle\ge0,
\qquad u\in\Dom\widetilde\Sigma.
\end{equation}
Assume that the common domain
\begin{equation}
\mathcal D:=\Dom H_W\cap\Dom\widetilde\Sigma
\end{equation}
is dense and that
$H_W-z-\widetilde\Sigma:\mathcal D\to\HH$
is bijective. Denote its everywhere-defined inverse by
$\widetilde R$. Suppose, in addition, that the Dyson identity
holds on the domain of the coefficient in the following weak sense:
\begin{equation}\label{eq:admissible-weak-dyson}
\langle\widetilde\Sigma u,f\rangle
=\sigma^2\int_{\Omega^c}
\langle\widetilde R v_a u,v_a f\rangle\dd a,
\qquad u\in\Dom\widetilde\Sigma,\quad f\in\HH.
\end{equation}
Then $\Dom\widetilde\Sigma=\HH$, the operator
$\widetilde\Sigma$ is bounded, and
\begin{equation}\label{eq:admissible-operator-bounds}
\|\widetilde R\|\le\frac1\tau,
\qquad \|\widetilde\Sigma\|\le\frac{\sigma^2\|v\|_{\HH}^2}{\tau}.
\end{equation}
Moreover,
\begin{equation}
\widetilde\Sigma=\Sigma_e(z),\qquad
\widetilde R=R_e(z),
\end{equation}
where the right-hand sides are the bounded-coefficient solution
and its resolvent constructed in Section~\ref{app:exist}.
The norms of $\widetilde R$ and $\widetilde\Sigma$ in
\eqref{eq:admissible-operator-bounds} are operator norms on $\HH$.
\end{lemma}

\begin{proof}
For $u\in\mathcal D$, self-adjointness of $H_W$ and
\eqref{eq:admissible-sign} give
\begin{equation}
-\im\langle(H_W-z-\widetilde\Sigma)u,u\rangle
=\tau\|u\|_{\HH}^2+\im\langle\widetilde\Sigma u,u\rangle
\ge\tau\|u\|_{\HH}^2.
\end{equation}
Cauchy--Schwarz therefore implies
\begin{equation}
\|(H_W-z-\widetilde\Sigma)u\|_{\HH}\ge\tau\|u\|_{\HH}.
\end{equation}
Taking $u=\widetilde R f$ for arbitrary $f\in\HH$ proves
$\|\widetilde R\|\le1/\tau$. Thus the inverse assumed to exist
in the lemma is automatically bounded.

For every $u\in\HH$, Tonelli's theorem gives
\begin{equation}
\int_{\Omega^c}\|v_a u\|_{\HH}^2\dd a
\le\int_{\mathbb R^3}\|v_a u\|_{\HH}^2\dd a
=\|v\|_{\HH}^2\|u\|_{\HH}^2.
\end{equation}
Consequently, for $u,f\in\HH$,
\begin{equation}
\sigma^2\int_{\Omega^c}
|\langle\widetilde R v_a u,v_a f\rangle|\dd a
\le \sigma^2\|\widetilde R\|
\left(\int_{\Omega^c}\|v_a u\|_{\HH}^2\dd a\right)^{1/2}
\left(\int_{\Omega^c}\|v_a f\|_{\HH}^2\dd a\right)^{1/2}
\le\frac{\sigma^2\|v\|_{\HH}^2}{\tau}\|u\|_{\HH}\|f\|_{\HH}.
\end{equation}
The integral in \eqref{eq:admissible-weak-dyson} is therefore
absolutely convergent and defines the bounded operator
\begin{equation}
T:=\Se[\widetilde R],\qquad
\|T\|\le\frac{\sigma^2\|v\|_{\HH}^2}{\tau}.
\end{equation}
Equation~\eqref{eq:admissible-weak-dyson}, with arbitrary $f$,
says exactly that $\widetilde\Sigma u=Tu$ for
$u\in\Dom\widetilde\Sigma$.

We now use closedness of $\widetilde\Sigma$ to identify its full
domain. Given any $u\in\HH$, choose
$u_n\in\Dom\widetilde\Sigma$ with $u_n\to u$ in $\HH$.
Boundedness of $T$ gives
$\widetilde\Sigma u_n=Tu_n\to Tu$.
Closedness means that convergence of both $u_n$ and
$\widetilde\Sigma u_n$ forces $u\in\Dom\widetilde\Sigma$
and $\widetilde\Sigma u=Tu$. Thus
$\Dom\widetilde\Sigma=\HH$ and $\widetilde\Sigma=T$.
This proves the second estimate in
\eqref{eq:admissible-operator-bounds}.

Now $\mathcal D=\Dom H_W$, and
\eqref{eq:admissible-sign} is precisely
$\im\widetilde\Sigma\ge0$ for this bounded operator.
Equation~\eqref{eq:admissible-weak-dyson} becomes the bounded
operator Dyson equation. The uniqueness theorem in
Section~\ref{app:exist} identifies $\widetilde\Sigma$ with
$\Sigma_e(z)$ and consequently identifies their resolvents.
\end{proof}

\subsubsection{A factorization with no inverse of \texorpdfstring{$Q$}{Q}}
\label{app:balanced}\label{sub:balanced-factor}
For fixed $z=E+i\tau$ with $\tau>0$, write $\Sigma_e=\Sigma_e(z)$ for the
exterior solution constructed in Section~\ref{app:exist}.
That section proves the properties used here:
\begin{equation}
\Sigma_e\in\BB(\HH),\qquad
\im\Sigma_e\ge0,\qquad
\|\Sigma_e\|\le \sigma^2\|v\|_{\HH}^2/\tau.
\end{equation}
These are conclusions of the construction, not additional assumptions.
The real and imaginary parts mean
\begin{equation}
\re\Sigma_e=\frac{\Sigma_e+\Sigma_e^*}{2},\qquad
\im\Sigma_e=\frac{\Sigma_e-\Sigma_e^*}{2i};
\end{equation}
both are bounded self-adjoint operators. We use the definitions of
$A$ and $\Gamma$ in \eqref{eq:factorization-data}.
Section~\ref{sub:fixed-point-construction}, Step~6, constructs the bounded solution $\Sigma_e(z)$
and proves $\|\Sigma_e(z)\|\le\sigma^2\|v\|_{\HH}^2/\tau$, independently of
Lemma~\ref{lem:mode-resolvent}. Consequently, $\Gamma$ is bounded
for every fixed $\tau>0$.

The balanced factorization used below is a standard device in Dyson-equation
analysis; see, for example, Ajanki--Erd\H{o}s--Kr\H{u}ger~\cite{AjEK}
and Alt--Erd\H{o}s--Kr\H{u}ger~\cite{AEK}. The point of the following
lemma is to establish the form needed here for the unbounded
Schr\H{o}dinger operator.

\begin{samepage}
\begin{lemma}[Factorization of the exterior resolvent]
\label{lem:exterior-factorization}
Under the hypotheses of Theorem~\ref{thm:general-exterior}, fix
$z=E+i\tau$ with $\tau>0$, and let $\Sigma_e=\Sigma_e(z)$ be the
self-energy constructed in Section~\ref{app:exist}.
Define $A$ and $\Gamma$ by \eqref{eq:factorization-data}.
The operator $T$ with the domain specified in
\eqref{eq:factorization-T} is self-adjoint, and the exterior resolvent
has the factorization
\begin{equation}\label{eq:operator-factor}
R_e=\Gamma^{-1/2}(T-i)^{-1}\Gamma^{-1/2}.
\end{equation}
The operators $D,Q,U$ in \eqref{eq:balanced-def} are bounded.
The operator $D$ is nonnegative
and injective, and
\begin{equation}\label{eq:balanced}
R_e=QUQ^*,\quad U^*U=UU^*=I,\quad
\im U=D>0,\quad Q^*\Gamma Q=D.
\end{equation}
Moreover,
\begin{equation}\label{eq:ImR}
\im R_e=QDQ^*.
\end{equation}
Here $D>0$ means nonnegative and injective.
\end{lemma}
\end{samepage}

\begingroup
\clubpenalty=10000
\widowpenalty=10000
\noindent{\em Remark.}
No bounded inverse of $D$ or $Q$ is required.
All boundedness assertions in this lemma concern fixed $\tau>0$;
no global bound uniform in $\tau$ for $Q$ or $\Gamma^{-1/2}$ is claimed.
No sign condition on $\re\Sigma_e$, spatial derivatives of $\Sigma_e$,
or multiplication-operator structure of $\Sigma_e$ is required.\par
\endgroup

\begin{proof}
The operator $A$ is self-adjoint on $\Dom H_W$, since
$-E-\re\Sigma_e$ is a bounded self-adjoint perturbation.
Here $H_W$ is the self-adjoint closure defined in
Section~\ref{sec:notation-assumptions};
$A$ need not be semibounded.
Since $\Gamma\ge\tau I$, the spectral theorem gives its bounded
inverse and $\|\Gamma^{-1}\|\le1/\tau$.

The boundedly invertible map $\Gamma^{-1/2}$ is a homeomorphism
of $\HH$, so the inverse image of the dense subspace $\Dom A$
is dense. Also $T$ is symmetric, because $A=A^*$ and
$\Gamma^{-1/2}$ is bounded and self-adjoint.

We verify that $T$ is self-adjoint, including its domain.
Let $y\in\Dom T^*$ and put $g=T^*y$. Then
\begin{equation}
\langle Tu,y\rangle=\langle u,g\rangle
\quad(u\in\Dom T).
\end{equation}
By \eqref{eq:factorization-T}, every such $u$ has the form $u=\Gamma^{1/2}h$ with $h\in\Dom A$.
Substituting $u=\Gamma^{1/2}h$ and using
$T=\Gamma^{-1/2}A\Gamma^{-1/2}$, we obtain
\begin{equation}
\langle Ah,\Gamma^{-1/2}y\rangle
=\langle h,\Gamma^{1/2}g\rangle
\quad(h\in\Dom A).
\end{equation}
By the definition of the adjoint, $\Gamma^{-1/2}y\in\Dom A^*$
and $A^*\Gamma^{-1/2}y=\Gamma^{1/2}g$.
Since $A^*=A$, this says $y\in\Dom T$ and $Ty=g$.
Thus $T^*\subset T$; symmetry gives the reverse inclusion.
This proves $T=T^*$.

On $\Dom A=\Dom H_W$ we have the exact factorization
\begin{equation}
A-i\Gamma=\Gamma^{1/2}(T-i)\Gamma^{1/2}.
\end{equation}
Its domain is exactly $\Dom A$: the condition
$\Gamma^{1/2}u\in\Dom T$ is equivalent to $u\in\Dom A$.
Since $T$ is self-adjoint, $T-i$ has a bounded inverse.
For $f\in\HH$, the vector
$u=\Gamma^{-1/2}(T-i)^{-1}\Gamma^{-1/2}f$ lies in $\Dom A$
by \eqref{eq:factorization-T} and satisfies $(A-i\Gamma)u=f$.
Since $R_e=(A-i\Gamma)^{-1}$, this proves
\eqref{eq:operator-factor}.

The scalar functions
\begin{equation}
t\longmapsto(1+t^2)^{-1/2},\qquad
t\longmapsto\frac{t+i}{\sqrt{1+t^2}}
\end{equation}
are bounded on $\mathbb R$. The first is strictly positive,
and the second has modulus one and imaginary part
$(1+t^2)^{-1/2}$. The spectral calculus therefore proves that
$D$ is bounded, nonnegative, and injective, that $U$ is unitary,
and that $\im U=D$. It also gives
\begin{equation}
D^{1/2}UD^{1/2}=(T-i)^{-1}.
\end{equation}
Substituting this identity into \eqref{eq:operator-factor} gives
$R_e=QUQ^*$. Boundedness of $Q$ follows from boundedness of
$\Gamma^{-1/2}$ and $D^{1/2}$. Direct multiplication gives
$Q^*\Gamma Q=D$.
Finally, taking imaginary parts of $R_e=QUQ^*$ yields
$\im R_e=Q(\im U)Q^*=QDQ^*$, proving \eqref{eq:ImR}.
\end{proof}

\section{Appendices}\label{sec:appendices}
\subsection{The momentum representation of the homogeneous covariance map}
\label{app:homogeneous-multiplier}

We prove \eqref{eq:momentum-convolution}. Let $f\in L^\infty(\mathbb R^3)$
and set $A_f=\mathcal F^{-1}M_f\mathcal F$.
Apply \eqref{eq:covariance-map-definition} with $A=\mathbb R^3$ and $T=A_f$.
Pairing the operator identity with Schwartz functions $u,w$ in physical space gives
\begin{equation}
\langle\SSS[A_f]u,w\rangle
=\sigma^2\int_{\mathbb R^3}\langle A_f(v_a u),v_a w\rangle\dd a,
\qquad (v_a u)(x)=v(x-a)u(x).
\end{equation}
Fix the center $a$. Since $\mathcal F A_f=M_f\mathcal F$, Plancherel's identity
in the spatial variable $x$ gives
\begin{equation}
\int_{\mathbb R^3}[A_f(v_a u)](x)\,\overline{(v_a w)(x)}\dd x
=\int_{\mathbb R^3}f(\eta)\widehat{v_a u}(\eta)\,
\overline{\widehat{v_a w}(\eta)}\dd\eta.
\end{equation}
Substituting the momentum integral for the inner product in the covariance formula
and integrating first in $a$, we obtain
\begin{equation}
\langle\SSS[A_f]u,w\rangle=\sigma^2\int_{\mathbb R^3}f(\eta)I(\eta)\dd\eta,
\qquad
I(\eta):=\int_{\mathbb R^3}\widehat{v_a u}(\eta)\,
\overline{\widehat{v_a w}(\eta)}\dd a.
\end{equation}

To evaluate $I(\eta)$, expand the two spatial Fourier transforms. This gives
\begin{equation}
I(\eta)=\frac{1}{(2\pi)^3}\iint e^{-i\eta\cdot(x-y)}
u(x)\overline{w(y)}
\left(\int v(x-a)v(y-a)\dd a\right)\dd x\dd y.
\end{equation}
Fourier inversion gives
\begin{equation}
\int v(x-a)v(y-a)\dd a
=\frac{1}{(2\pi)^3}\int e^{ip\cdot(x-y)}|\widehat v(p)|^2\dd p.
\end{equation}
Substituting this expression into $I(\eta)$ combines the two exponential
factors into $e^{-i(\eta-p)\cdot(x-y)}$. The $x$- and $y$-integrals
then give $(2\pi)^3\widehat u(\eta-p)\overline{\widehat w(\eta-p)}$.
With $\xi=\eta-p$, we obtain
\begin{equation}
I(\eta)=\frac{1}{(2\pi)^3}\int_{\mathbb R^3}|\widehat v(\eta-\xi)|^2
  \widehat u(\xi)\overline{\widehat w(\xi)}\dd\xi.
\end{equation}
Substituting this formula for $I(\eta)$ into $\langle\SSS[A_f]u,w\rangle$
and interchanging the $\xi$- and $\eta$-integrals gives the multiplier
of $\SSS[A_f]$. Since $v$ is real, $|\widehat v(\eta-\xi)|=|\widehat v(\xi-\eta)|$.
Now \eqref{eq:momentum-convolution} follows for every $u\in\HH$.

\subsection{Verification of the transformed covariance identity}
\label{app:transformed-covariance-identity}

We verify \eqref{eq:transformed-self-consistency}. By the definition
\eqref{eq:saturated} and the identity $QDQ^*=\im R_e$ from
\eqref{eq:ImR},
\begin{equation}
\Phi_e[D]
=Q^*\Se[QDQ^*]Q
=Q^*\Se[\im R_e]Q.
\end{equation}
The covariance map $\Se$ is complex-linear and preserves adjoints.
Taking imaginary parts of the Dyson equation $\Sigma_e=\Se[R_e]$
therefore gives
\begin{equation}
\im\Sigma_e
=\frac{\Se[R_e]-\Se[R_e]^*}{2i}
=\Se\!\left[\frac{R_e-R_e^*}{2i}\right]
=\Se[\im R_e].
\end{equation}
Hence $\Phi_e[D]=Q^*(\im\Sigma_e)Q$.

Next, recall $\Gamma=\tau I+\im\Sigma_e$ and
$Q^*\Gamma Q=D$ from \eqref{eq:balanced}. Thus
\begin{equation}
Q^*(\im\Sigma_e)Q=Q^*(\Gamma-\tau I)Q=Q^*\Gamma Q-\tau Q^*Q=D-\tau Q^*Q.
\end{equation}
Combining the preceding identities proves
\eqref{eq:transformed-self-consistency}.

\subsection{From local estimates to the global operator bound}
\label{app:local-to-global}

We give the details of the deduction of \eqref{eq:general-coefficient-bound} from the
local estimate \eqref{eq:K-local-coef}. Fix $z\in\mathbb C_+$ and
suppress it in the notation. For $f\in\HH$, let
$f_p=\one_{C_p}f$, where $C_p=p+[-1/2,1/2)^3$, $p\in\mathbb Z^3$.
Since the cubes form a disjoint partition of $\mathbb R^3$,
\begin{equation}\label{eq:local-global-partition}
f=\sum_p f_p\quad\text{in }\HH,
\qquad
\sum_p\|f_p\|_{\HH}^2=\|f\|_{\HH}^2.
\end{equation}

The function $\beta_{C_p}(a)=\sup_{x\in C_p}v(x-a)$ is bounded by
$\|v\|_\infty$ and vanishes unless $a$ lies within distance $s$
of $C_p$. This neighborhood is contained in a cube of side $1+2s$.
Therefore
\begin{equation}\label{eq:local-global-A-bound}
A_{C_p}^2=\int_{\Omega^c}\beta_{C_p}(a)\dd a
\le\|v\|_\infty(1+2s)^3.
\end{equation}
Set $K_v^2=\sigma^2\|v\|_\infty^2(1+2s)^3$.
By \eqref{eq:K-local-coef} and \eqref{eq:local-global-A-bound},
$\|\Sigma_e\one_{C_p}\|\le K_v$ for every $p$.
Applying this operator bound to $f_p=\one_{C_p}f_p$ gives
\begin{equation}\label{eq:local-global-piece-bound}
\|\Sigma_e f_p\|_{\HH}^2
=\|(\Sigma_e\one_{C_p})f_p\|_{\HH}^2
\le K_v^2\|f_p\|_{\HH}^2.
\end{equation}
The norm on the right is the norm of the individual piece $f_p$;
its square can therefore be summed using
\eqref{eq:local-global-partition}.

We next use the finite support of $v$ to control the overlap of the
functions $\Sigma_e f_p$. In the integral formula
$\Sigma_e f_p=\sigma^2\int_{\Omega^c}v_aR_ev_af_p\dd a$,
the rightmost factor $v_af_p$ vanishes unless the center $a$ is
within distance $s$ of $C_p$. The leftmost factor $v_a$ restricts
the output to points within distance $s$ of $a$. Consequently,
$\Sigma_e f_p$ vanishes almost everywhere outside
$E_p=\{x\in\mathbb R^3:\dist(x,C_p)\le2s\}$.
The number of these enlarged cubes containing any fixed point is bounded:
\begin{equation}\label{eq:local-global-overlap}
N_s:=\sup_{x\in\mathbb R^3}\#\{p\in\mathbb Z^3:x\in E_p\}
\le C(1+s)^3,
\end{equation}
where $C$ is a numerical constant. Indeed, if $x\in E_p$, then
each coordinate of the center $p$ differs from the corresponding
coordinate of $x$ by at most $2s+1/2$, leaving at most $C(1+s)^3$
possible lattice points $p$.

First suppose that only finitely many $f_p$ are nonzero. At almost every
point $x$, at most $N_s$ terms in $\sum_p(\Sigma_e f_p)(x)$ can be
nonzero. The Cauchy--Schwarz inequality thus gives
\begin{equation}
\left|\sum_p(\Sigma_e f_p)(x)\right|^2
\le N_s\sum_p|(\Sigma_e f_p)(x)|^2.
\end{equation}
Integrating in $x$, applying \eqref{eq:local-global-piece-bound}, and
then using \eqref{eq:local-global-partition}, we obtain
\begin{equation}\label{eq:local-global-summation}
\begin{aligned}
\|\Sigma_e f\|_{\HH}^2
&=\left\|\sum_p\Sigma_e f_p\right\|_{\HH}^2
\le N_s\sum_p\|\Sigma_e f_p\|_{\HH}^2\\
&\le N_sK_v^2\sum_p\|f_p\|_{\HH}^2
=N_sK_v^2\|f\|_{\HH}^2.
\end{aligned}
\end{equation}

For arbitrary $f\in\HH$, let $f^{(n)}$ be the sum of $f_p$ over
$p\in\mathbb Z^3$ with $\max_j|p_j|\le n$.
Then $f^{(n)}\to f$ in $\HH$ by \eqref{eq:local-global-partition}.
The boundedness of $\Sigma_e$ for each fixed $z$, already established
in Section~\ref{app:exist}, gives
$\Sigma_e f^{(n)}\to\Sigma_e f$ in $\HH$.
Passing to the limit in \eqref{eq:local-global-summation}, applied
to $f^{(n)}$, proves that inequality for every $f\in\HH$.
Taking square roots and the supremum over $\|f\|_{\HH}=1$ yields
\begin{equation}\label{eq:local-global-operator-bound}
\|\Sigma_e\|\le\sqrt{N_s}\,K_v=:C_v.
\end{equation}
This is \eqref{eq:general-coefficient-bound}. The constant depends only on $v$, $\sigma^2$,
and $s$, and is independent of $\Omega$, $W$, and $z$.
The same argument applies to $\Sigma_e^*$, using the adjoint estimate
in \eqref{eq:K-local-coef} and replacing $R_e$ by $R_e^*$ in the
integral representation.

\phantomsection

\end{document}